\documentclass[final,review,3p,times]{elsarticle}

\usepackage{amssymb}
\usepackage{amsmath}
\usepackage{hyperref}
\usepackage{graphbox}
\usepackage{orcidlink}
\usepackage{circuitikz}
\usepackage{calc}
\usepackage{subcaption}
\usetikzlibrary{calc}
\usetikzlibrary{patterns}
\usetikzlibrary{arrows.meta}
\usepackage[export]{adjustbox}
\usepackage[nameinlink,noabbrev,capitalize]{cleveref}
\usepackage{subcaption}
\usepackage{multirow}
\usepackage{booktabs}
\usepackage{tabularx}
\hypersetup{colorlinks=true, linkcolor=blue,urlcolor=blue,filecolor=blue,citecolor=red,allcolors=blue}
\usepackage{bm}
\usepackage{soul}
\usepackage{listings}
\usepackage{xcolor}
\usepackage{pdfpages}

\definecolor{codegreen}{rgb}{0,0.6,0}
\definecolor{codegray}{rgb}{0.5,0.5,0.5}
\definecolor{codepurple}{rgb}{0.58,0,0.82}
\definecolor{backcolour}{rgb}{0.95,0.95,0.92}

\lstdefinestyle{mystyle}{
    backgroundcolor=\color{backcolour},   
    commentstyle=\color{codegreen},
    keywordstyle=\color{magenta},
    numberstyle=\tiny\color{codegray},
    stringstyle=\color{codepurple},
    basicstyle=\ttfamily\footnotesize,
    breakatwhitespace=false,         
    breaklines=true,                 
    captionpos=b,                    
    keepspaces=true,                 
    numbers=left,                    
    numbersep=5pt,                  
    showspaces=false,                
    showstringspaces=false,
    showtabs=false,                  
    tabsize=2
}
\usepackage{lineno}

\journal{\href{https://www.sciencedirect.com/journal/european-journal-of-mechanics-a-solids}{European Journal of Mechanics - A/Solids}}

\begin{document}
\includepdf[pages=1]{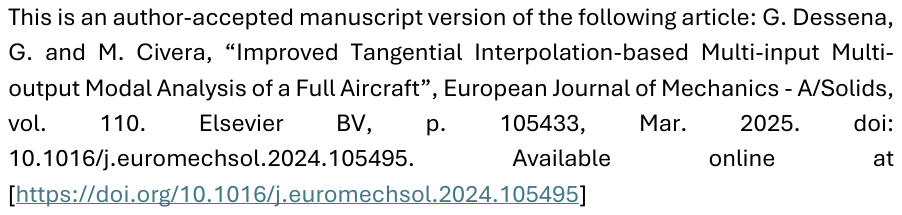}
\begin{frontmatter}


\title{Improved Tangential Interpolation-based Multi-input Multi-output Modal Analysis of a Full Aircraft\tnoteref{label5}}
\tnotetext[label5]{\hl{This is an author-accepted manuscript version of the following article: G. Dessena and M. Civera, “Improved Tangential Interpolation-based Multi-input Multi-output Modal Analysis of a Full Aircraft”, European Journal of Mechanics - A/Solids, vol. 110. Elsevier BV, p. 105495, Mar. 2025. doi: 10.1016/j.euromechsol.2024.105495. Available online at} [\url{https://doi.org/10.1016/j.euromechsol.2024.105495}]}

\author[label1]{Gabriele Dessena \orcidlink{0000-0001-7394-9303}}
\affiliation[label1]{organization={Department of Aerospace Engineering, Universidad Carlos III de Madrid},
             addressline={Av.da de la Universidad 30},
             postcode={28911},
             city={Leganés},
             state={Madrid},
             country={Spain}}

\author[label2]{Marco Civera \orcidlink{0000-0003-0414-7440}}

\affiliation[label2]{organization={Department of Structural, Building and Geotechnical Engineering, Politecnico di Torino},
            addressline={Corso Duca degli Abruzzi 24}, 
            city={Turin},
            postcode={10129}, 
            state={Piedmont},
            country={Italy}}

\begin{abstract}
In the field of Structural Dynamics, modal analysis is the foundation of System Identification and vibration-based inspection. However, despite their widespread use, current state-of-the-art methods for extracting modal parameters from multi-input multi-output (MIMO) frequency domain data are still affected by many technical limitations. Mainly, they can be computationally cumbersome and/or negatively affected by close-in-frequency modes. The Loewner Framework (LF) was recently proposed to alleviate these problems with the limitation of working with single-input data only. This work proposes a computationally improved version of the {LF}, or iLF, to extract modal parameters more efficiently. Also, the proposed implementation is extended in order to handle MIMO data in the frequency domain. This new implementation is compared to state-of-the-art methods such as the frequency domain implementations of the Least Square Complex Exponential method and the Numerical Algorithm for Subspace State Space System Identification on numerical and experimental datasets. More specifically, a finite element model of a 3D Euler-Bernoulli beam is used for the baseline comparison and the noise robustness verification of the proposed MIMO iLF algorithm. Then, an experimental dataset from MIMO ground vibration tests of a trainer jet aircraft with over 91 accelerometer channels is chosen for the algorithm validation on a real-life application. {Its validation is carried out with known results from a single-input multi-output dataset of the starboard wing of the same aircraft}. Excellent results are achieved in terms of accuracy, robustness to noise, and computational performance by the proposed improved MIMO method, both on the numerical and the experimental datasets. The MIMO iLF MATLAB implementation is shared in the work supplementary material. 
\end{abstract}


\begin{highlights}
\item {A computationally improved version of the Loewner Framework (iLF) is proposed;}
\item {Multi-input multi-output (MIMO) modal parameters extraction via the iLF is achieved;}
\item {The iLF is more accurate and computationally efficient than previous implementations;}
\item {The proposed method outperforms well-established methods, such as N4SID and LSCE;}
\item {The iLF is used for the modal identification of a BAE Systems Hawk T1A aircraft from MIMO data.}
\end{highlights}


\end{frontmatter}


\section{Introduction}
Modal analysis plays an important role during the design and validation of industrial systems and products \cite{Zheng2023}. In particular, the goal of Experimental Modal Analysis (EMA) is to extract the modal parameters -- natural frequencies ($\omega_n$), damping ratios ($\zeta_n$), and mode shapes ($\mathbf{\phi}_n$) -- of the system under analysis \cite{Kumar2022}. These are generally employed for finite element model updating \cite{Dessena2022c} or vibration-based Structural Health Monitoring (SHM) \cite{Civera2021a}. 

The modal parameters of the systems under scrutiny are extracted through System Identification (SI) techniques, which are usually defined according to the type of data they are capable of processing. First, some techniques are meant for output-only data, where the input is unknown a priori, and others for input-output data. The former is the foundation of operational modal analysis and lies outside the scope of this work. The latter, known as EMA, represents the context of this work. EMA methods can be further divided with respect to the number of input and output channels, considering one or more of them. Single-input single-output (SISO) and single-input multi-output (SIMO) modal analysis are the most widely used methods for EMA as they are easier and cheaper to implement than their counterparts, especially multi-input multi-output (MIMO). That derives from the need {for} more advanced equipment and dedicated algorithms. In most cases, SIMO testing can suffice the scope of the test, such as when only one in-plane direction needs to be investigated \cite{Uwayed2023}. However, at other times, a MIMO approach might be needed to fully characterise some of the modes of the system, such as in \cite{Dessena2022b}, where an out-of-plane mode is not fully characterised in the SIMO test.  

In addition, despite SI being a mature and well-explored field \cite{Ljung2020}, other-than-commercial openly available implementations of MIMO methods for modal analysis are very scarce. This is another hurdle for the implementation of more MIMO-based testing. Nevertheless, the MIMO algorithms have the same drawbacks as the existing SIMO methods: the ill-conditioning of the fitting matrix resulting in high computational cost and time \cite{Civera2021}. Hence, a technique based on rational interpolation along tangential directions, the Loewner Framework (LF), has been extended for the extraction of the modal parameters from mechanical systems by the authors \cite{Dessena2022,Dessena2022f,Dessena2022g}. 

{Thus}, this work aims to introduce a novel computationally improved version of the LF (iLF) to extend the modal parameters extraction capability of the LF from SIMO to full MIMO capability. {The main innovative aspects are not limited to the extension from SIMO to MIMO datasets; as it will be discussed in further detail, the newly introduced version of the LF, iLF, is designed to be computationally more efficient than the traditional LF. This is a significant advancement, as the original LF suffered from computational bottlenecks, especially when handling large datasets or high-dimensional systems.}

To achieve this aim, this work tackles the following:
\begin{enumerate}
    \item in \cref{sec:met} (Methodology), the key aspects of MIMO and the conventional LF approach are recalled; the proposed MIMO iLF implementation is then fully detailed;
    \item in \cref{sec:num} (Numerical study), the proposed iLF approach is firstly validated on numerical data from a simulated structure, with a very simplified geometry (cantilever hollow beam). This allows to investigate the potential of the procedure in the most controlled fashion possible;
    \item again in \cref{sec:num}, the newly-proposed iLF is compared to the standard LF approach, introduced in previous works \cite{Dessena2022}, as well as to other state-of-the-art approaches - namely, the Least Square Complex Exponential (LSCE) \cite{Zimmerman2017} and the frequency domain numerical algorithm for subspace state space system identification (N4SID) \cite{VanOverschee2005}. Both accuracy and computational requirements are benchmarked for all procedures. The effects of artificially added white Gaussian noise (WGN) are thoroughly investigated as well to assess the effectiveness with increasingly noisy measurements;
    \item The proposed methodology is then experimentally validated in \cref{sec:exp}. Data from controlled laboratory Ground Vibration Tests on a full-scale advanced training jet aircraft are used, both from a single wing and from the whole structure. As before, the results from other different classic algorithms are reported for direct comparability;
    \item The Conclusions (\cref{sec:conc}) ends this paper.
\end{enumerate}

\section{Methodology}\label{sec:met}
\subsection{Multi-input multi-output testing for modal parameters extraction}
As {already hinted before}, MIMO testing is an advanced method in structural dynamics and modal analysis for experimentally extracting modal parameters from complex systems. Unlike traditional SISO and SIMO methods, MIMO testing involves simultaneously applying multiple inputs and measuring multiple responses. These multiple inputs might be in different positions on the specimen or excite the structure in different directions.

Two state-of-the-art techniques for the extraction of modal parameters from MIMO data are LSCE \cite{Zimmerman2017} and N4SID \cite{VanOverschee2005}. they are, respectively, based on the least-squares solution of the complex exponential formulation and stochastic subspace identification. In this work, their frequency domain MATLAB implementations, functions \verb|modalfit|\footnote{\url{https://uk.mathworks.com/help/signal/ref/modalfit.html}} and \verb|n4sid|\footnote{\url{https://uk.mathworks.com/help/ident/ref/n4sid.html}}, are used solely for benchmarking purposes.

A few examples of MIMO tests can be found in the current scientific literature, with different applications. In \cite{Milewicz2022}, an inner disc and railway wheel rim are tested using multiple impact hammers as input. Notably, \cite{Aglietti2019}, while discussing the potential of digital twins for virtual testing, introduces the HYDRA test facility at the European Space Research and Technology Centre, which includes a 6 degrees-of-freedom vibration table for full-scale satellites. In aeronautics, \cite{Peres2014} discusses the MIMO test, using two sting shakers as input, of a helicopter rotor spider and the advantages of the MIMO setup compared to a SISO setup for the characterisation of the system. 

The ultimate goal of MIMO in modal analysis is to obtain a complete characterisation of the system
in a single test. This is what, to be validated, the newly proposed technique has to obtain in a computationally efficient manner.

\subsection{The Loewner Framework}

The conventional LF algorithm has been applied to {the modelling of a multi-port electrical system} \cite{Lefteriu2009} {and to aerodynamic model order reduction for aeroservoelastic modelling in} \cite{Quero2019}. {The authors introduced LF for the identification of modal parameters from mechanical systems in} \cite{Dessena2022}{, validated its computational performance in }\cite{Dessena2022f} {and its noise robustness characteristic for SHM in }\cite{Dessena2022g}{. However, these studies were limited in their applicability to SIMO systems only.} This Section first outlines the general background of the LF and then expands on the computational improvements brought by the iLF {and its extension to MIMO systems}. 

Lets begin by defining the Loewner matrix $\boldsymbol{\mathbb{L}}$:
\noindent \emph{Given a row array of pairs of complex numbers ($\mu_j$,${v}_j$), $j=1$,...,$q$, and a column array of pairs of complex numbers ($\lambda_i$,${w}_j$), $i=1$,...,$k$, with $\lambda_i$, $\mu_j$ distinct, the associated $\boldsymbol{\mathbb{L}}$, or divided-differences matrix is:}
\begin{equation}
\label{eq:LM}
\boldsymbol{\mathbb{L}}=\begin{bmatrix}
\frac{\mathbf{v}_1-\mathbf{w}_1}{\mu_1-\lambda_1} & \cdots & \frac{\mathbf{v}_1-\mathbf{w}_k}{\mu_1-\lambda_k}\\
\vdots & \ddots & \vdots\\
\frac{\mathbf{v}_q-\mathbf{w}_1}{\mu_q-\lambda_1} & \cdots & \frac{\mathbf{v}_q-\mathbf{w}_k}{\mu_q-\lambda_k}\\
\end{bmatrix}\:\in \mathbb{C}^{q\times k}
\end{equation}
\emph{If there is a known underlying function $\pmb{\phi}$, then $\mathbf{w}_i=\pmb{\phi}(\lambda_i)$ and $\mathbf{v}_j=\pmb{\phi}(\mu_j).$}

Karl Löwner established a connection between $\boldsymbol{\mathbb{L}}$ and rational interpolation, also known as Cauchy interpolation \cite{Lowner1934}. This allows to define interpolants based on determinants of submatrices of $\boldsymbol{\mathbb{L}}$. According to~\cite{Antoulas2017,Mayo2007}, rational interpolants  can be derived from $\boldsymbol{\mathbb{L}}$. The approach based on the Loewner pencil is considered in this work. The Loewner pencil consists of the $\boldsymbol{\mathbb{L}}$ and $\boldsymbol{\mathbb{L}}_s$ matrices, where $\boldsymbol{\mathbb{L}}_s$ is the \emph{Shifted Loewner matrix}, defined later.

In order to {describe} how the LF works, let us consider a linear time-invariant dynamical system $\mathbf{\Sigma}$ with $m$ inputs and $p$ outputs, and $k$ internal variables in descriptor-form representation, given by:
\begin{equation}
\mathbf{\Sigma}:\;\mathbf{E}\frac{d}{dt}\mathbf{x}(t)=\mathbf{A}\mathbf{x}(t)+\mathbf{B}\mathbf{u}(t);\;\;\;
\mathbf{y}(t)=\mathbf{C}\mathbf{x}(t)+\mathbf{D}\mathbf{u}(t)
 \label{eq:LTI}
 \end{equation}

\noindent where $\mathbf{x}(t)\:\in\: \mathbb{R}^{k}$ is the internal variable, $\mathbf{u}(t)\:\in\:\mathbb{R}^{m}$ is the function's input and $\mathbf{y}(t)\:\in\:\mathbb{R}^{p}$ is the output. The constant system matrices are:
\begin{equation}
    \mathbf{E},\mathbf{A}\in \mathbb{R}^{k\times k},\; \mathbf{B}\in \mathbb{R}^{k\times m}\; \mathbf{C}\in \mathbb{R}^{p\times k}\; \mathbf{D}\in \mathbb{R}^{p\times m}
\end{equation}

\noindent a Laplace transfer function, $\mathbf{H}(s)$, of $\mathbf{\Sigma}$ can be defined in the form of a $p\times m$ rational matrix function, when the matrix $\mathbf{A}-\lambda\mathbf{E}$ is non singular for a given finite value $\lambda$, such that $\lambda \in \mathbb{C}$:
\begin{equation}
    \mathbf{H}(s)=\mathbf{C}(s\mathbf{E}-\mathbf{A})^{-1}\mathbf{B}+\mathbf{D}
    \label{eq:trans}
\end{equation}

\noindent Let us consider the more general case of tangential interpolation (rational interpolation along tangential directions~\cite{Kramer2016}). The right interpolation data becomes:

\begin{equation}
\begin{gathered}
   (\lambda_i;\mathbf{r}_i,\mathbf{w}_i),\: i = 1,\dots,\rho
    \quad \quad
    \begin{matrix}
        \mathbf{\Lambda}=\text{diag}[\lambda_1,\dotsc,\lambda_k]\in \mathbb{C}^{\rho\times \rho}\\
        \mathbf{R}=[\mathbf{r}_1\;\dotsc \mathbf{r}_k]\in \mathbb{C}^{m\times \rho}\\
        \mathbf{W} = [\mathbf{w}_1\;\dotsc\;\mathbf{w}_k]\in \mathbb{C}^{\rho\times \rho}
        \end{matrix}\Bigg\}
\end{gathered}
\label{eq:RID}
\end{equation}
 
 \noindent Likewise, the left interpolation data:
\begin{equation}
    \begin{gathered} 
        (\mu_j,\mathbf{l}_j,\mathbf{v}_j),\: j = 1,\dots,v
        \quad \quad
        \begin{matrix}
            \mathbf{M}=\mathrm{diag}[\mu_1,\dotsc,\mu_q]\in \mathbb{C}^{v\times v}\\
            \mathbf{L}^T=[\mathbf{l}_1\;\dotsc \mathbf{l}_v]\in \mathbb{C}^{p\times v}\\
            \mathbf{V}^T = [\mathbf{v}_1\;\dotsc\;\mathbf{v}_q]\in \mathbb{C}^{m\times v}
        \end{matrix}\Bigg\}
    \end{gathered}
\label{eq:LID}
\end{equation}

\noindent $\lambda_i$ and $\mu_j$ are the values at which $\mathbf{H}(s)$ is evaluated (the frequency bins in this application). The vectors $\mathbf{r}_i$ and $\mathbf{l}_j$ are, respectively, the right and left tangential general directions, randomly selected in practice~\cite{Quero2019}, and $\mathbf{w}_i$ and $\mathbf{v}_j$ are the right and left tangential data. Linking $\mathbf{w}_i$ and $\mathbf{v}_j$ to the transfer function $\mathbf{H}$, associated with realisation $\mathbf{\Sigma}$ in \cref{eq:LTI}, solves the rational interpolation problem:

\begin{equation}
 \begin{split}
\mathbf{H}(\lambda_i)\mathbf{r}_i=\mathbf{w}_i,\:j=1.\dots,\rho \quad \text{and} \quad
\mathbf{l}_i\mathbf{H}(\mu_j)=\mathbf{v}_j,\:i=1,\dots,v
 \end{split}
 \label{eq:LS2}
 \end{equation}
\noindent such that the Loewner pencil satisfies \cref{eq:LS2}.

Now, let us consider a set of points $Z=\{z_1,\dots,z_N\}$ in the complex plane, a rational function $\mathbf{y}(s)$, such that $\mathbf{y}_i:=\mathbf{y}(z_i),i=1,\dots,N$, and let $Y=\{\mathbf{y}_1,\dots,\mathbf{y}_N\}$. Including the left and right data partitions, the following is obtained:

\begin{equation}
\begin{split}
	Z=\{\lambda_1,\dots,\lambda_\rho\} \cup \{\mu_1,\dots,\mu_v\}\quad \text{and} \quad
	Y=\{\mathbf{w}_1,\dots,\mathbf{w}_\rho\} \cup \{\mathbf{v}_1,\dots,\mathbf{v}_v\}
\end{split}
\label{eq:ZY}
\end{equation}
with $N=p+v$. Thus, the matrix $\boldsymbol{\mathbb{L}}$ becomes:

\begin{equation}
\label{eq:LM2}
\boldsymbol{\mathbb{L}}=\begin{bmatrix}
\frac{\mathbf{v}_1\mathbf{r}_1-\mathbf{l}_1\mathbf{w}_1}{\mu_1-\lambda_1} & \cdots & \frac{\mathbf{v}_1\mathbf{r}\rho-\mathbf{l}_1\mathbf{w}\rho}{\mu_1-\lambda\rho}\\
\vdots & \ddots & \vdots\\
\frac{\mathbf{v}_v\mathbf{r}_1-\mathbf{l}_v\mathbf{w}_1}{\mu_v-\lambda_1}& \cdots & \frac{\mathbf{v}_v\mathbf{r}\rho-\mathbf{l}_v\mathbf{w}\rho}{\mu_v-\lambda\rho}\\
\end{bmatrix}\:\in \mathbb{C}^{v\times \rho}
\end{equation}

Since $\mathbf{v}_v\mathbf{r}_p$ and $\mathbf{l}_v\mathbf{w}_p$ are scalars, the Sylvester equation is satisfied by $\boldsymbol{\mathbb{L}}$ as follows:
\begin{equation}
    \boldsymbol{\mathbb{L}}\mathbf{\Lambda}-\mathbf{M}\boldsymbol{\mathbb{L}}=\mathbf{L}\mathbf{W}-\mathbf{V}\mathbf{R}
    \label{eq:syl}
\end{equation}

Now, Let us define the \emph{shifted Loewner matrix}, $\boldsymbol{\mathbb{L}}_s$, as the $\boldsymbol{\mathbb{L}}$ corresponding to $s\mathbf{H}(s)$:

\begin{equation}
\label{eq:LS}
\boldsymbol{\mathbb{L}}_s=\begin{bmatrix}
\frac{\mu_1\mathbf{v}_1\mathbf{r}_1-\lambda_1\mathbf{l}_1\mathbf{w}_1}{\mu_1-\lambda_1} & \cdots & \frac{\mu_1\mathbf{v}_1\mathbf{r}_\rho-\lambda_\rho\mathbf{l}_1\mathbf{w}_\rho}{\mu_1-\lambda_\rho}\\
\vdots & \ddots & \vdots\\
\frac{\mu_v\mathbf{v}_v\mathbf{r}_1-\lambda_1\mathbf{l}_v\mathbf{w}_1}{\mu_v-\lambda_1}& \cdots & \frac{\mathbf{v}_v\mathbf{r}_\rho-\mathbf{l}_v\mathbf{w}_\rho}{\mu_v-\lambda_\rho}\\
\end{bmatrix}\:\in \mathbb{C}^{v\times \rho}
\end{equation}

Likewise, the Sylvester equation is satisfied as follows:
\begin{equation}
    \boldsymbol{\mathbb{L}}_s\Lambda-\mathbf{M}\boldsymbol{\mathbb{L}}_s=\mathbf{L}\mathbf{W}\mathbf{\Lambda}-\mathbf{M}\mathbf{V}\mathbf{R}
    \label{eq:syl2}
\end{equation}

Let us focus on {considering} \cref{eq:trans} and matrix $\mathbf{D}$. As shown in~\cite{Mayo2007}, $\mathbf{D}$ is set to $0$ per convention as its contributions are incorporated in the other matrices. So, \cref{eq:trans} becomes:

\begin{equation}
\mathbf{H}(s)=\mathbf{C}(s\mathbf{E}-\mathbf{A})^{-1}\mathbf{B}
\label{eq:fin}
\end{equation}

A realisation with the smallest possible dimension exists only if the system is fully controllable and observable. Therefore, assuming the data is sampled from a system whose transfer function is characterised by \cref{eq:fin}, the generalised tangential observability, $\mathcal{O}v$, and generalised tangential controllability, $\mathcal{R}\rho$, are defined as follows~\cite{Lefteriu2010b}:
\begin{equation}
    \mathcal{O}_v=\begin{bmatrix}
    \mathbf{l}_1\mathbf{C}(\mu_1\mathbf{E}-\mathbf{A})^{-1}\\
    \vdots\\
   \mathbf{l}_v\mathbf{C}(\mu_q\mathbf{E}-\mathbf{A})^{-1}
    \end{bmatrix}\:\in \mathbb{R}^{v\times n}
    \label{eq:Oq}
\end{equation}
\begin{equation}
    \mathcal{R}_\rho=\begin{bmatrix}
    (\lambda_1\mathbf{E}-\mathbf{A})^{-1}\mathbf{Br}_1 &
    \cdots &
    (\lambda_\rho\mathbf{E}-\mathbf{A})^{-1}\mathbf{Br}_\rho 
    \end{bmatrix}\:\in \mathbb{R}^{n\times \rho}
    \label{eq:Rk}
\end{equation}

Now, let us incorporate \cref{eq:Oq,eq:Rk} into, respectively, \cref{eq:LM2,eq:LS}:

\begin{equation}
\begin{split}
    \boldsymbol{\mathbb{L}}_{j,i} = \dfrac{\mathbf{v}_j\mathbf{r}_i-\mathbf{l}_j\mathbf{w}_i}{\mu_j-\lambda_i}=\dfrac{\mathbf{l}_j\mathbf{H}(\mu_i)\mathbf{r}_i-\mathbf{l}_j\mathbf{H}(\lambda_i)\mathbf{r}_i}{\mu_j-\lambda_i}=
    -\mathbf{l}_j\mathbf{C}(\mu_j\mathbf{E-\mathbf{A}})^{-1}\mathbf{E}(\lambda_i\mathbf{E}-\mathbf{A})^{-1}\mathbf{Br}_i
\end{split}
\label{eq:LH}
\end{equation}

\begin{equation}
\begin{split}
    (\boldsymbol{\mathbb{L}}_s)_{j,i} = \dfrac{\mu_j\mathbf{v}_j-\lambda_i\mathbf{l}\mathbf{w}_i}{\mu_j-\lambda_i}
    =\dfrac{\mu_j\mathbf{l}_j\mathbf{H}(\mu_i)\mathbf{r}_i-\lambda_i\mathbf{l}_j\mathbf{H}(\lambda_i)\mathbf{r}_i}{\mu_j-\lambda_i}=\\= -\mathbf{l}_j\mathbf{C}(\mu_j\mathbf{E}-\mathbf{A})^{-1}\mathbf{A}(\lambda_i\mathbf{E}-\mathbf{A})^{-1}\mathbf{Br}_i
\end{split}
\label{eq:lLH}
\end{equation}

Firstly, let us consider the case with a minimal amount of data, where it is assumed that $p = v$. This assumption is based on the fact that no duplicate data is permitted in $\mathbf{R}$ and $\mathbf{L}$. {Thus, rearranging} \cref{eq:Oq,eq:Rk} into \labelcref{eq:LH} and \labelcref{eq:lLH}:
\begin{align}
    \boldsymbol{\mathbb{L}}=-\mathcal{O}_v\mathbf{E}\mathcal{R}_\rho &&
    \boldsymbol{\mathbb{L}}_s=-\mathcal{O}_v\mathbf{A}\mathcal{R}_\rho
    \label{eq:LLs}
\end{align}

Then, letting the Loewner pencil be a regular pencil, in the sense of $\mathrm{eig}((\boldsymbol{\mathbb{L}},\boldsymbol{\mathbb{L}}_s))\neq (\mu_i,\lambda_i)$:
\begin{align}
    \mathbf{E}=-\boldsymbol{\mathbb{L}},&&\mathbf{A}=-\boldsymbol{\mathbb{L}}_s,&&\mathbf{B}=\mathbf{V},&&\mathbf{C}=\mathbf{W}
\end{align}
Accordingly, the interpolating rational function is defined as follows:
\begin{equation}
    \mathbf{H}(s)=\mathbf{W}(\boldsymbol{\mathbb{L}}_s-s\boldsymbol{\mathbb{L}})^{-1}\mathbf{V}
\end{equation}
The derivation presented applies to the minimal data scenario, a case rarely encountered in real-world datasets. Nevertheless, the LF can be expanded to handle redundant data points effectively. 

Firstly, let us assume:

\begin{equation}
\begin{split}
    \mathrm{rank}[\zeta\boldsymbol{\mathbb{L}}-\boldsymbol{\mathbb{L}}_s]=\mathrm{rank}[\boldsymbol{\mathbb{L}}\:\boldsymbol{\mathbb{L}}_s]=
   =\mathrm{rank}
    \begin{bmatrix}
    \boldsymbol{\mathbb{L}}\\ \boldsymbol{\mathbb{L}}_s
    \end{bmatrix}=k,\; 
    \forall \zeta \in \{\lambda_j\}\cup\{\mu_i\}
    \end{split}
    \label{eq:cond1}
\end{equation}
Secondly, the short Singular Value Decomposition (SVD) of $\zeta\boldsymbol{\mathbb{L}}-\boldsymbol{\mathbb{L}}_s$ is computed:
\begin{equation}
    \textrm{svd}(\zeta\boldsymbol{\mathbb{L}}-\boldsymbol{\mathbb{L}}_s)=\mathbf{Y}\mathbf{\Sigma}_l\mathbf{X}
\label{eq:cond2}
\end{equation}
where $\mathrm{rank}(\zeta\boldsymbol{\mathbb{L}}-\boldsymbol{\mathbb{L}}_s)=\mathrm{rank}(\mathbf{\Sigma}_l)=\mathrm{size}(\mathbf{\Sigma}_l)=k,\mathbf{Y}\in\mathbb{C}^{v \times k}$ and $\mathbf{X}\in\mathbb{C}^{k\times \rho}$.
Thirdly, note that:
\begin{equation}
\begin{split}
-\mathbf{A}\mathbf{X}+\mathbf{E}\mathbf{X}\mathbf{\Sigma}_l = \mathbf{Y}^*\boldsymbol{\mathbb{L}}_s\mathbf{X}^*\mathbf{X}-\mathbf{Y}^*\boldsymbol{\mathbb{L}}\mathbf{X}^*\mathbf{X}\mathbf{\Sigma}_l=\mathbf{Y}^*(\boldsymbol{\mathbb{L}}_s-\boldsymbol{\mathbb{L}}\mathbf{\Sigma}_l )=\mathbf{Y}^*\mathbf{V}\mathbf{R}=\mathbf{BR}
\end{split}
\label{eq:cond3}
\end{equation}
and likewise, $-\mathbf{Y}\mathbf{A}+\mathbf{M}\mathbf{Y}\mathbf{E}=\mathbf{L}\mathbf{C}$ such that $\mathbf{X}$ and $\mathbf{Y}$ are, respectively, the generalised controllability and observability matrices for the system $\mathbf{\Sigma}$ with $\mathbf{D}=0$. 
After verifying the right and left interpolation conditions, the Loewner realisation for redundant data is given by:
\begin{equation}
\begin{split}
\mathbf{E}=-\mathbf{Y}^*\boldsymbol{\mathbb{L}}\mathbf{X},\quad \quad
\mathbf{A}=-\mathbf{Y}^*\boldsymbol{\mathbb{L}}_s\mathbf{X,}\quad \quad
\mathbf{B}=\mathbf{Y}^*\mathbf{V},\quad \quad
\mathbf{C}=\mathbf{W}\mathbf{X}
\end{split}
\label{eq:real}
\end{equation}

The formulation of \cref{eq:real} — that is, the Loewner realisation for redundant data — will be employed throughout the remainder of this work. For a more detailed discussion of each step, the interested reader is referred to \cite{Mayo2007,Antoulas2017}. Finally, the modal parameters of the system can then be extracted by eigenanalysis of the system matrices $\mathbf{A}$ and $\mathbf{C}$ in \cref{eq:real}.

\subsubsection{Computational improvements}
As briefly mentioned in the earlier Sections, this work introduces an improved version of the LF. In this iLF implementation, the computational inefficiencies of the original implementation are amended to increase performance and {the implementation issues resolved to allow for MIMO modal identification capabilities.} 

In terms of inefficiencies in the original LF implementation for the extraction of mechanical systems, the following are identified:
\begin{enumerate}
   \item \textbf{The \texttt{for} loops in the LF implementation originally reported in} \cite{Lefteriu2009,Lefteriu2010a,Ionita}\textbf{:} {The most computationally cumbersome part was the coordinate transformation to obtain matrices with real entries from the LF. This was an iterative step in the LF original implementation, once available in }\cite{Ionita}\footnote{At the time of writing, it is no longer possible to download the original implementation from the original source, previously found at \url{http://aci.rice.edu/system-identification/}. An implementation, modified by the first author, is available at \url{https://dspace.lib.cranfield.ac.uk/handle/1826/22465}.}.{ This is amended in the iLF with a vectorised approach based on the Kronecker tensor product }(\verb|kron|\footnote{\url{https://uk.mathworks.com/help/control/ref/kron.html}}{ in MATLAB). The code snippets of the former and current implementations are found in} \hyperref[sec:app]{Appendix A.1}.
   \item \textbf{The way the matrices $\mathbf{A}$ and $\mathbf{E}$ are handled in the LF for modal parameters extraction in} \cite{Dessena2022,Dessena2021}\textbf{:} {In the authors' original contribution for modal parameters extraction with LF,  a descriptor state-space model obtained from LF was converted to a continuous state-space model with the }\verb|dss|\footnote{\url{https://uk.mathworks.com/help/control/ref/dss.html}} 
 and \verb|ss|\footnote{\url{https://uk.mathworks.com/help/control/ref/ss.html}} {MATLAB functions. This was a considerable bottleneck. Now amended with simple matrices operations to obtain the continuous state-space matrix} $\mathbf{A}$ for the extraction of modal parameters. See \hyperref[sec:app2]{Appendix A.2} {for the MATLAB code snippets of the original and improved implementation.}
\end{enumerate}

{With respect to the capability expansion from SIMO to MIMO, the second improvement in the code allowed for a slick transformation also of multi-input data. With the old implementation, this was a cumbersome and computationally demanding process, which limited the LF to SIMO data. The iLF implementation is openly available in}  \cite{Dessena2024e}.

\section{Numerical study}\label{sec:num}
\begin{figure}[!ht]
\centering
	\begin{subfigure}[t]{.4\textwidth}
	\centering
		{\includegraphics[align=c,width=\textwidth,keepaspectratio]{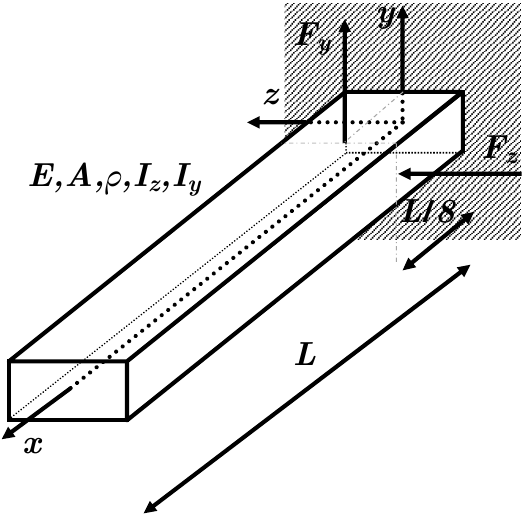}}
		\captionsetup{font={it},justification=centering}
		\subcaption{\label{fig:3dbeam}}	
	\end{subfigure}
    \begin{subfigure}[t]{.45\textwidth}
	\centering
		\includegraphics[align=c,width=\textwidth,keepaspectratio]{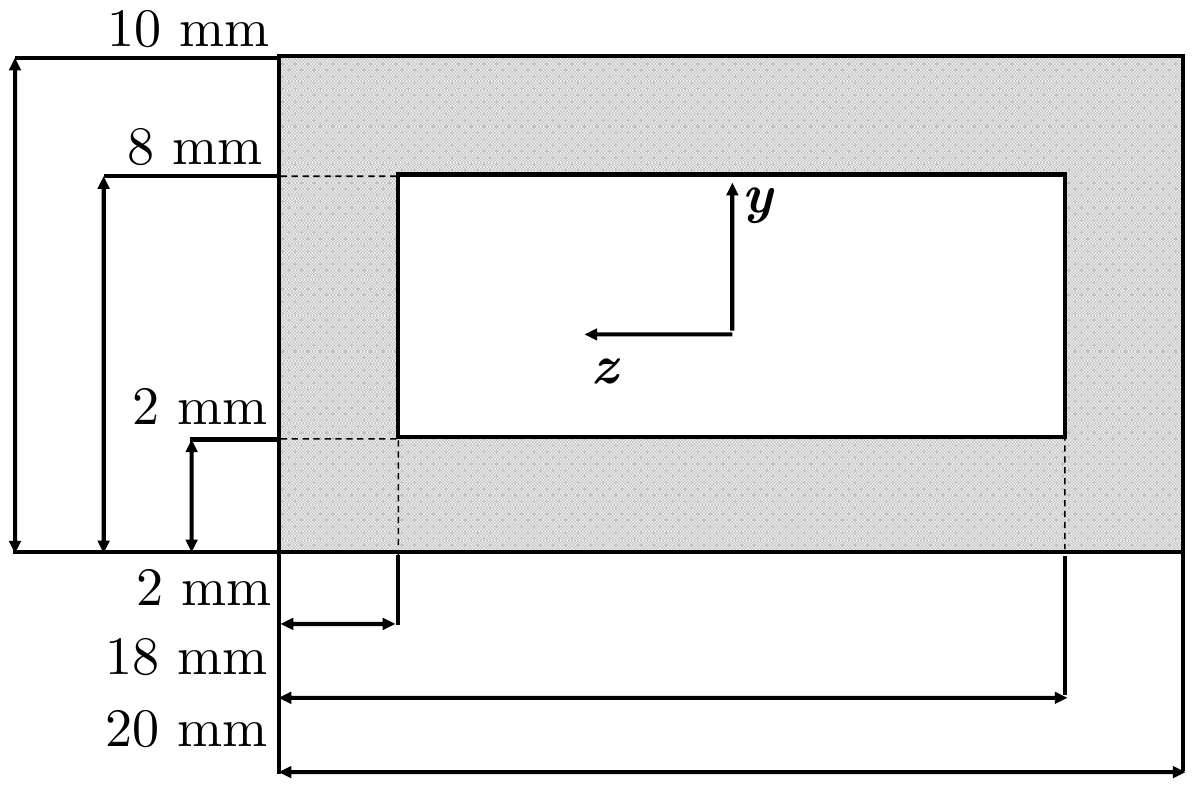}
		\captionsetup{font={it},justification=centering}
		\subcaption{\label{fig:beam_sec}}	
	\end{subfigure}
	\caption{8-element 3D beam: \cref{fig:3dbeam} is the schematic drawing of the discretised 8-element Euler-Bernoulli beam and \cref{fig:beam_sec} shows the beam rectangular box cross-section with the relative dimensions. Not in scale.}
	\label{fig:beam_n}
\end{figure}

In this Section, a 3D Euler Bernoulli beam is introduced to validate the newly proposed iLF MIMO implementation. The beam is modelled as a cantilever beam discretised in 8 elements and 1.5 m long (L). The beam material is aluminium, with a Young modulus (E) of 69 GPa and a density ($\rho$) of 2700 kgm\textsuperscript{-3}. The beam cross-section is shown in \cref{fig:beam_sec}, alongside the beam discretisation (\cref{fig:3dbeam}).
As already mentioned earlier, this very simple {finite element model} is intended as a controlled environment {to evaluate the capabilities of the proposed MIMO iLF approach,} especially in terms of robustness to artificially-introduced measurement noise. Field test investigation on a real-life case study will follow in the next Section.

The excitation of the numerically simulated beam is applied at node \# 2 (at a distance L/8 from the root) with a 1 N step input in the y and z directions. Rotations and x-axial displacement responses are ignored, resulting, therefore, in 16 non-zero displacement time series (in m; please note that the root node is constrained at the wall) and 2 input force time histories (in N, corresponding to the y- and z-direction input). All the input and output signals are sampled at 800 Hz, allowing inspections up to 400 Hz due to the Nyquist criterion. A damping ratio of 3\% is arbitrarily set for all modes. The MIMO Compliance FRFs ($\mathbf{H}(\omega)$) of the system take the shape of \verb|m:n:p| (MATLAB notation), where \verb|m| is the number of outputs, \verb|n| is the number of inputs, and \verb|p| is the number of frequency bins. They are obtained as follows, where $\mathbf{Y}(\omega)$ and $\mathbf{X}(\omega)$ are, respectively, the frequency domain representations of the output displacements and input forces obtained through the Fast Fourier Transform:
\begin{equation}
    \mathbf{Y}(\omega) = \mathbf{H}(\omega)\mathbf{X}(\omega)
    \label{eq:9}
\end{equation}

The computed FRFs are then used to identify the modal parameters through the iLF, the standard LF, and the two benchmark methods -- N4SID and LSCE. {All these algorithms were implemented in MATLAB.}

\subsection{Modal parameters identification}

\begin{figure}[!ht]
\centering
	\begin{subfigure}[t]{.495\textwidth}
	\centering
		{\includegraphics[width=\textwidth,keepaspectratio]{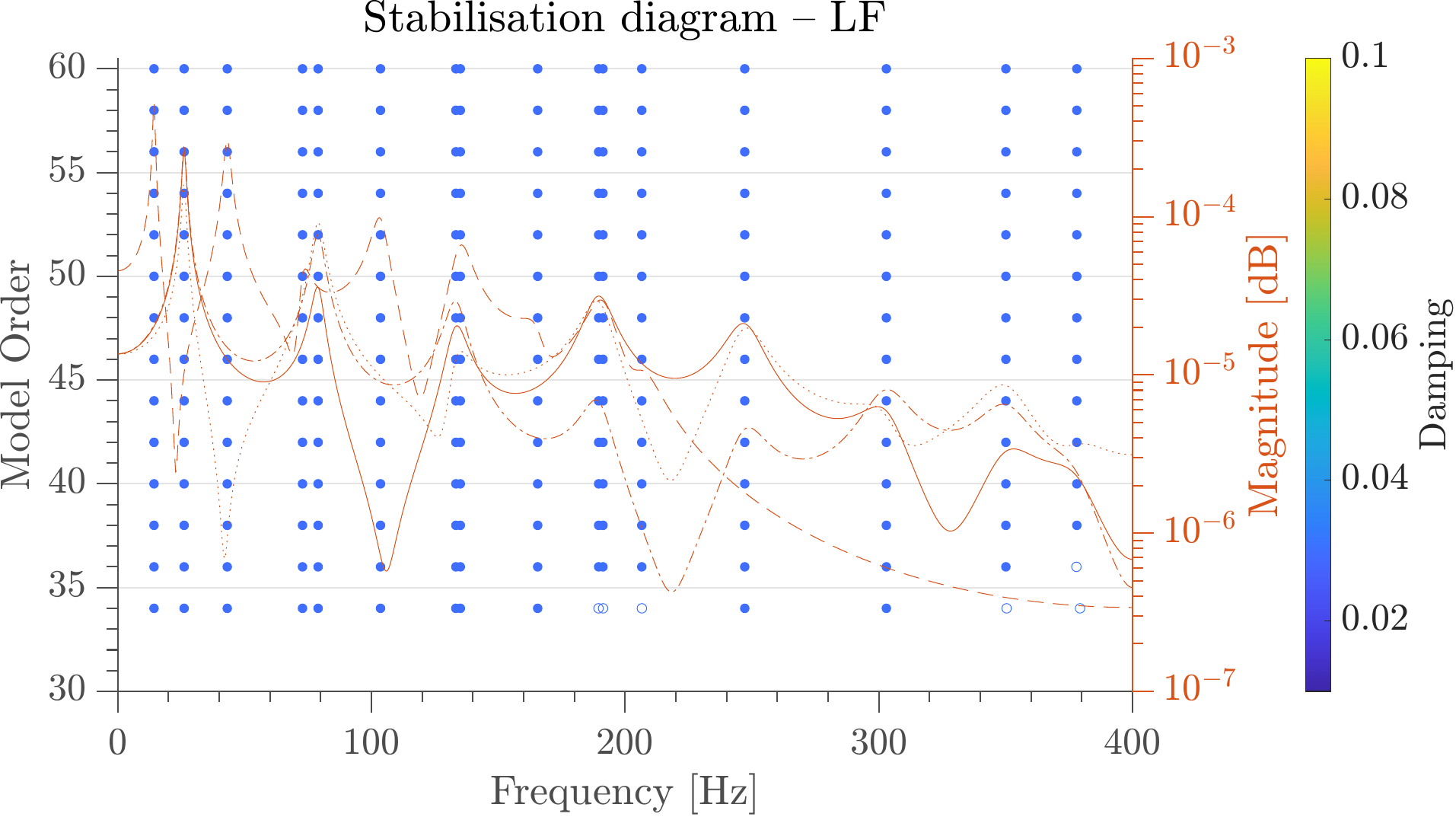}}
		\captionsetup{font={it},justification=centering}
		\subcaption{\label{fig:LF_stab}}	
	\end{subfigure}
    \begin{subfigure}[t]{.495\textwidth}
	\centering
		\includegraphics[width=\textwidth,keepaspectratio]{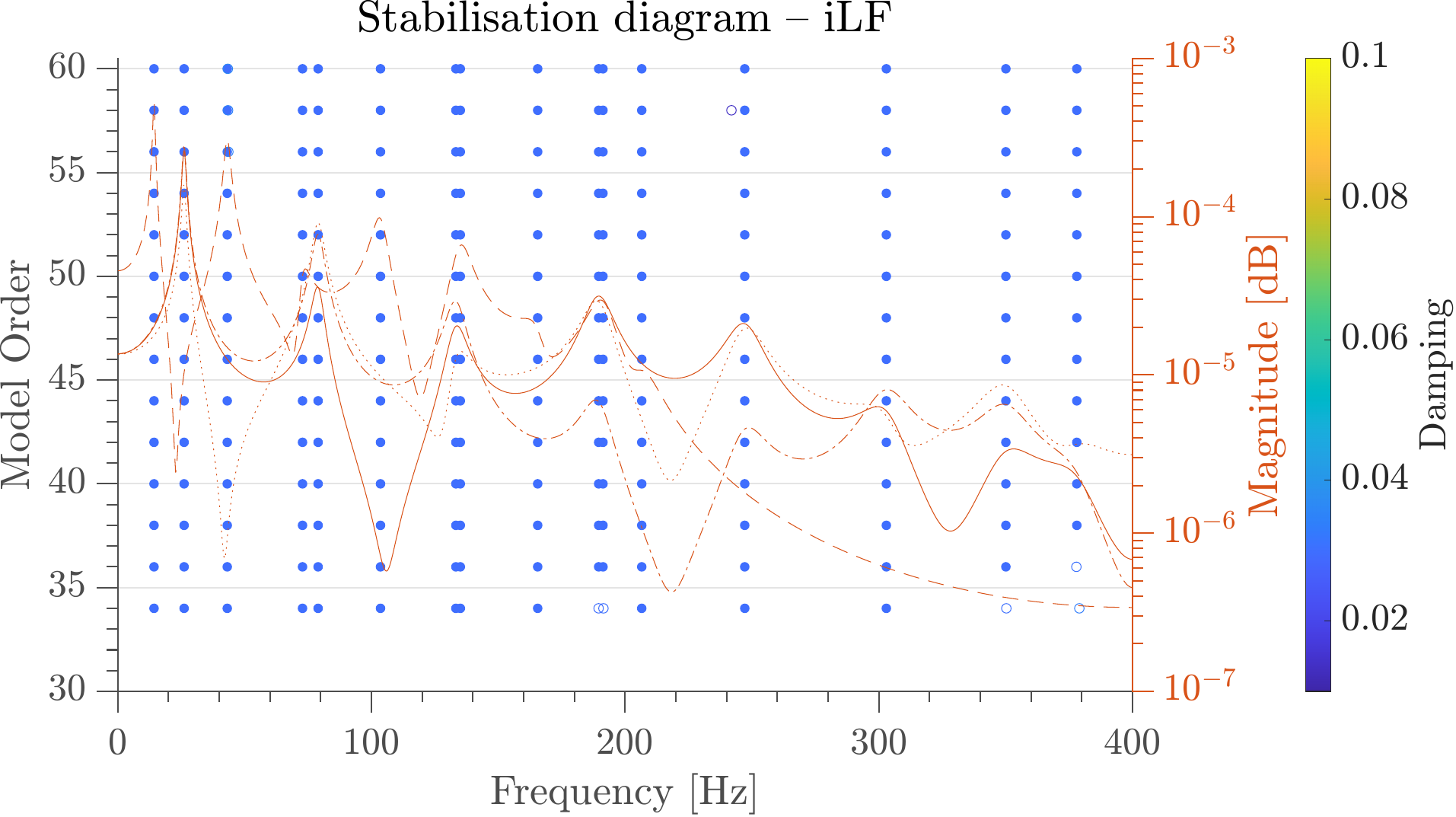}
		\captionsetup{font={it},justification=centering}
		\subcaption{\label{fig:iLF_stab}}	
	\end{subfigure}
    \begin{subfigure}[t]{.495\textwidth}
	\centering
		{\includegraphics[width=\textwidth,keepaspectratio]{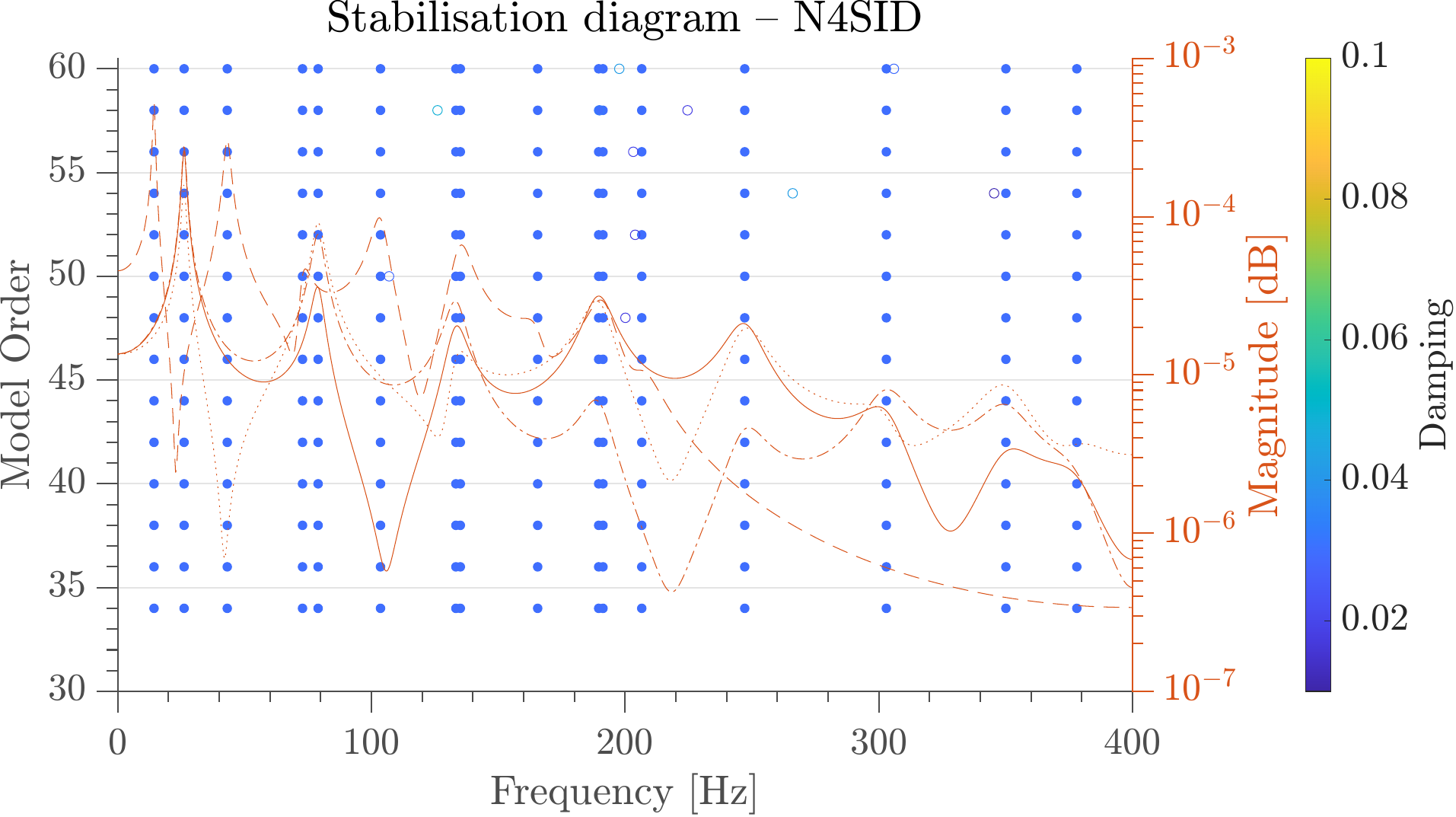}}
		\captionsetup{font={it},justification=centering}
		\subcaption{\label{fig:N4SID_stab}}	
	\end{subfigure}
    \begin{subfigure}[t]{.495\textwidth}
	\centering
		\includegraphics[width=\textwidth,keepaspectratio]{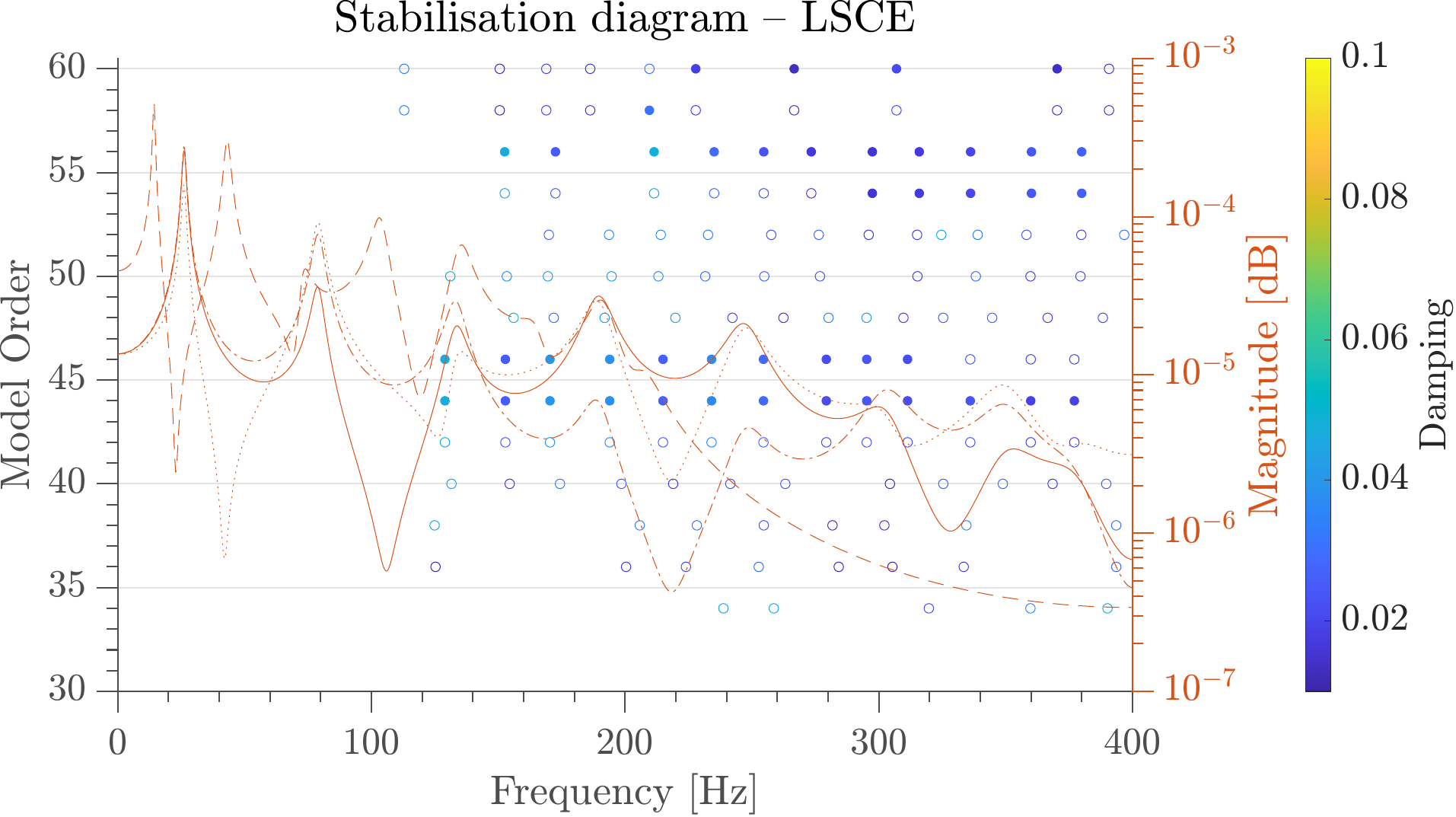}
		\captionsetup{font={it},justification=centering}
		\subcaption{\label{fig:LSCE_stab}}	
	\end{subfigure}
	\caption{8-element 3D beam: Stabilisation diagrams for the model parameters identification via LF (\cref{fig:LF_stab}), iLF (\cref{fig:iLF_stab}), N4SID (\cref{fig:N4SID_stab}), and LSCE (\cref{fig:LSCE_stab}).}
	\label{fig:beam_stab}
\end{figure}

The first step for assessing the suitability of iLF for the identification of modal parameters from MIMO measurements is to compare its performance. This is done here in terms of both precision and computational efficiency. Hence, the said FRFs were considered to identify the relevant modal parameters with N4SID, LSCE, LF (the standard implementation with the extra MIMO capability), and iLF to compare it with the analytical results from the 3D beam mass ($\mathbf{M}$) and stiffness ($\mathbf{K}$) matrices.
The modal parameters are obtained by iterating over different orders $k$ for the four methods and extracting the relative modes using stabilisation diagrams between orders 32 (minimum order of the system) and 60. The resulting stabilisation diagrams are shown in \cref{fig:beam_stab}, and the identified $\omega_n$ are reported in \cref{tab:beam_freq}.

From \cref{fig:beam_stab}, it is clear that all methods, LSCE apart, can identify 16 stable modes. The {relatively poor performance of} the LSCE {algorithm} is not unusual. In fact, in \cite{Dessena2022}, LSCE was found to struggle in situations where modes were close to each other. In the system under investigation, one can notice that modes \#4-5 and 7-8 are relatively close in frequency. On the computational side of things, it means that LSCE can only build rank-deficient matrices for approximating the system, hence misinterpreting it{, or partially identifying it}.

\begin{table}[!hbt]
\centering
\caption{Natural frequencies identified by MIMO iLF and the benchmark method. In parenthesis the relative error w.r.t. to the analytical result.\label{tab:beam_freq}}
\resizebox{.55\textwidth}{!}{
\begin{tabular}{lccccc}
\hline
\multicolumn{6}{c}{\textbf{Natural Frequency} [Hz] -- (Difference w.r.t. analytical) [\%]}                                                                                           \\ \hline
\textbf{Mode \#}            & \textbf{Analytical} & \textbf{N4SID}  & \textbf{LSCE} & \textbf{LF} & \textbf{iLF} \\ \hline
\multirow{2}{*}{\textbf{1}}  & 14.277  & 14.284  & -    & 14.284  & 14.284  \\
                             &         & (0.049) & (-)    & (0.049) & (0.049) \\
\multirow{2}{*}{\textbf{2}}  & 26.132  & 26.144  & -    & 26.144  & 26.144  \\
                             &         & (0.046) & (-)    & (0.046) & (0.046) \\
\multirow{2}{*}{\textbf{3}}  & 43.123  & 43.142  & -    & 43.142  & 43.142  \\
                             &         & (0.044) & (-)    & (0.044) & (0.044) \\
\multirow{2}{*}{\textbf{4}}  & 72.778  & 72.811  & -    & 72.811  & 72.811  \\
                             &         & (0.045) & (-)    & (0.045) & (0.045) \\
\multirow{2}{*}{\textbf{5}}  & 78.929  & 78.964  & -    & 78.964  & 78.964  \\
                             &         & (0.044) & (-)    & (0.044) & (0.044) \\
\multirow{2}{*}{\textbf{6}}  & 103.509 & 103.556 & -    & 103.556 & 103.556 \\
                             &         & (0.045) & (-)    & (0.045) & (0.045) \\
\multirow{2}{*}{\textbf{7}}  & 133.208 & 133.268 & -    & 133.268 & 133.268 \\
                             &         & (0.045) & (-)    & (0.045) & (0.045) \\
\multirow{2}{*}{\textbf{8}}  & 134.952 & 135.012 & -    & 135.012 & 135.012 \\
                             &         & (0.044) & (-)    & (0.044) & (0.044) \\
\multirow{2}{*}{\textbf{9}}  & 165.432 & 165.506 & -    & 165.506 & 165.506 \\
                             &         & (0.045) & (-)    & (0.045) & (0.045) \\
\multirow{2}{*}{\textbf{10}} & 189.455 & 189.541 & -    & 189.541 & 189.541 \\
                             &         & (0.045) & (-)    & (0.045) & (0.045) \\
\multirow{2}{*}{\textbf{11}} & 191.161 & 191.247 & -    & 191.247 & 191.247 \\
                             &         & (0.045) & (-)    & (0.045) & (0.045) \\
\multirow{2}{*}{\textbf{12}} & 206.433 & 206.526 & -    & 206.526 & 206.526 \\
                             &         & (0.045) & (-)    & (0.045) & (0.045) \\
\multirow{2}{*}{\textbf{13}} & 247.006 & 247.117 & 227.811 & 247.117 & 247.117 \\
                             &         & (0.045) & (-7.771) & (0.045) & (0.045) \\
\multirow{2}{*}{\textbf{14}} & 302.794 & 302.930 & 266.604 & 302.930 & 302.930 \\
                             &         & (0.045) & (-11.952) & (0.045) & (0.045) \\
\multirow{2}{*}{\textbf{15}} & 349.887 & 350.044 & 306.939 & 350.044 & 350.044 \\
                             &         & (0.045) & (-12.275) & (0.045) & (0.045) \\
\multirow{2}{*}{\textbf{16}} & 377.839 & 378.009 & 370.251 & 378.009 & 378.009 \\
                             &         & (0.045) & (-2.008)  & (0.045) & (0.045) \\
\hline
\end{tabular}}
\end{table}

Obtaining stable modes is not the only task required for an identification technique. Indeed, the identified modes need to be coherent with their real-life, or in this case analytical, counterparts. Frequency-wise, this can be assessed in terms of divergence between the identified $\omega_n$ and the known target values.

As anticipated, LSCE cannot obtain {the full set of modes of the system}. In particular, the LSCE identified {only the last four modes $\omega_n$, with absolute errors orders of magnitude higher than the other methods (the maximum absolute error is found for mode 16 and was over 12}\%). On the other hand, all other techniques showed similar results with an error, in terms of $\omega_n$, of less than 0.05\%. The same, even if not included in \cref{tab:beam_freq} for brevity, happens for $\zeta_n$, with the only exception of LSCE. For which errors over 29\% are found.

In terms of $\mathbf{\phi}_n$, the Modal Assurance Criterion (MAC) value \cite{Allemang1982} between the modes identified and the analytical ones is {(close to)} 1, exception made {for the modes LSCE could not identify}. \Cref{tab:mac_value} {shows the diagonal terms (off-diagonal terms are negligible) of the MAC values between the analytical} $\mathbf{\phi}_n$ {and those identified.}

\begin{table}[!hbt]
\centering
\caption{{The diagonal terms of the MAC values between the analytical} $\mathbf{\phi}_n$ {and those identified.} \label{tab:mac_value}}
\begin{tabular}{lcccc}
\hline
\textbf{Mode} & \textbf{N4SID} & \textbf{LSCE} & \textbf{LF} & \textbf{iLF} \\ \hline
\textbf{1}  & 0.915  & -      & 0.915  & 0.915  \\
\textbf{2}  & 0.974  & -      & 1      & 1      \\
\textbf{3}  & 1      & -      & 1      & 1      \\
\textbf{4}  & 1      & -      & 1      & 1      \\
\textbf{5}  & 1      & -      & 1      & 0.997  \\
\textbf{6}  & 0.998  & -      & 0.998  & 0.998  \\
\textbf{7}  & 0.999  & -      & 0.999  & 0.999  \\
\textbf{8}  & 0.999  & -      & 0.999  & 0.999  \\
\textbf{9}  & 0.999  & -      & 0.999  & 0.999  \\
\textbf{10} & 1      & -      & 0.999  & 1      \\
\textbf{11} & 1      & -      & 1      & 1      \\
\textbf{12} & 1      & -      & 1      & 1      \\
\textbf{13} & 1      & 1      & 1      & 1      \\
\textbf{14} & 1      & 1      & 1      & 1      \\
\textbf{15} & 1      & 1      & 1      & 1      \\
\textbf{16} & 1      & 1      & 1      & 1      \\
\hline
\end{tabular}
\end{table}

\subsection{Computational performance}

\begin{figure}[!ht]
	\centering
		{\includegraphics[align=c,width=.6\textwidth,keepaspectratio]{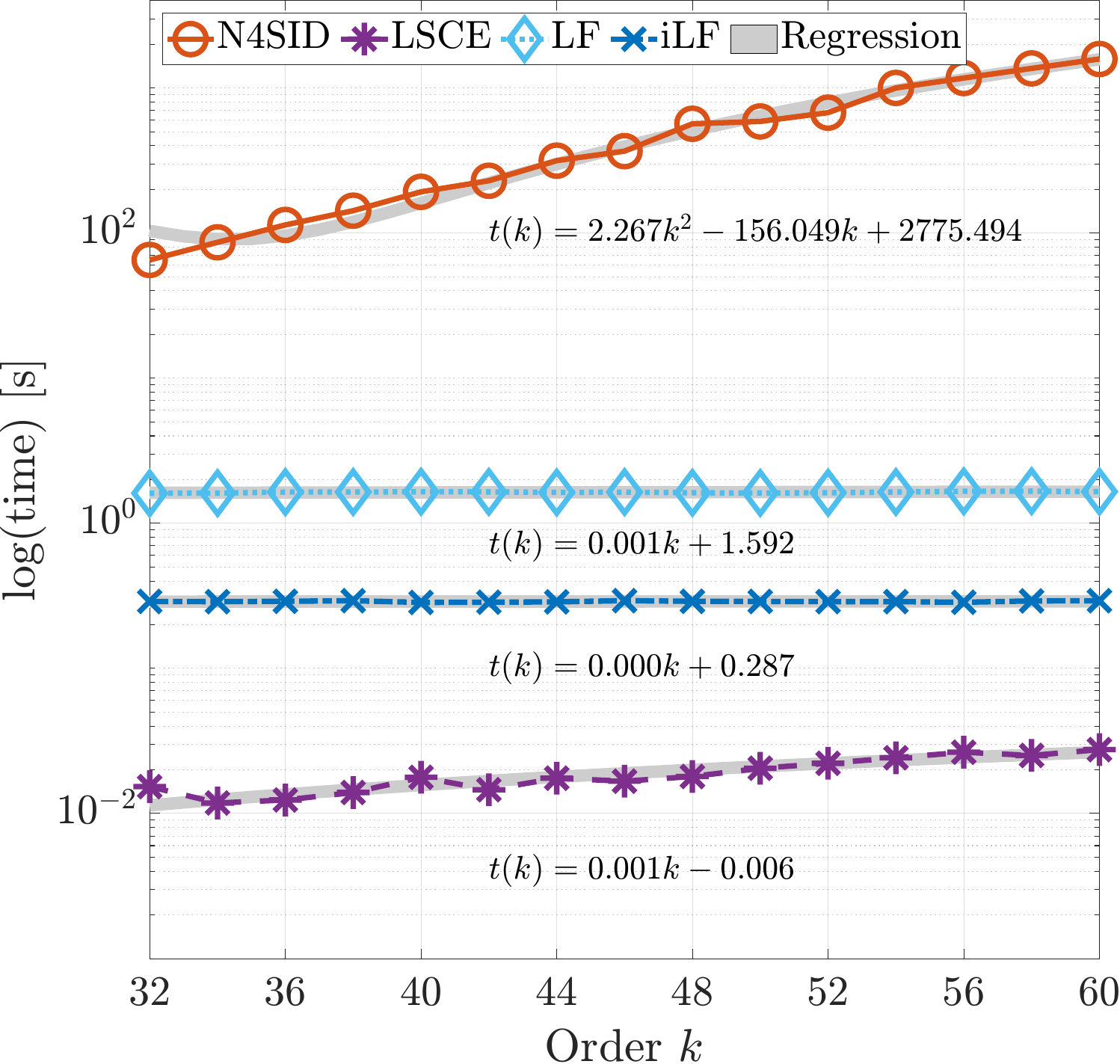}}
	\caption{8-element 3D beam: Average time to identification for all methods considered for orders $k$ between 32 and 60.}
	\label{fig:beam_time}
\end{figure}

The computational performance of the methods is assessed by measuring the time to identification at every order $k$ $\in$ [32, 60]. To achieve accurate results, each measurement is repeated ten times; hence, a mean and a standard deviation are available for each identification. In \cref{fig:beam_time}, the mean values are presented as solid points, and the standard deviation is neglected as it accounts for a negligible percentage of the mean value. The time to identification is measured on a 2019 Mac Pro base model with 64 GB of RAM running MATLAB 2021b.  

As already shown in \cite{Dessena2022f} for the SIMO modal parameters identification, N4SID is orders of magnitude slower than LF and LSCE. LSCE is found to be the fastest method, but it is clear that it does not serve a purpose as it cannot identify all modes. However, it's important to see that the iLF is an order of magnitude faster than the traditional implementation of LF. In addition, the conventional LF shows a slight dependency on model order, which is not the case for the iLF. The regressed line equation verifies this in \cref{fig:beam_time}.

\subsection{Investigation of noise effects}
After having preliminary validated the iLF identification against the traditional LF and the benchmark methods, the next step is to assess its robustness to noise. To achieve this, the time series are corrupted with different levels, with respect to the standard deviation of the signal, of additive WGN. The selected levels are 0.1, 0.5, 1, 1.5 and 2\% (note that both the input and output signals are corrupted with noise). Then, the corrupted time series are used to obtain the FRFs, which in turn are fed to iLF to obtain the modal parameters. The validation process is carried out in two steps: (i) the minimum order $k$ = 16 is considered, and (ii) the stable modes between orders 32 and 60 are examined.

\begin{figure*}[!ht]
\centering
	\begin{subfigure}[t]{.325\textwidth}
	\centering
		{\includegraphics[width=\textwidth,keepaspectratio]{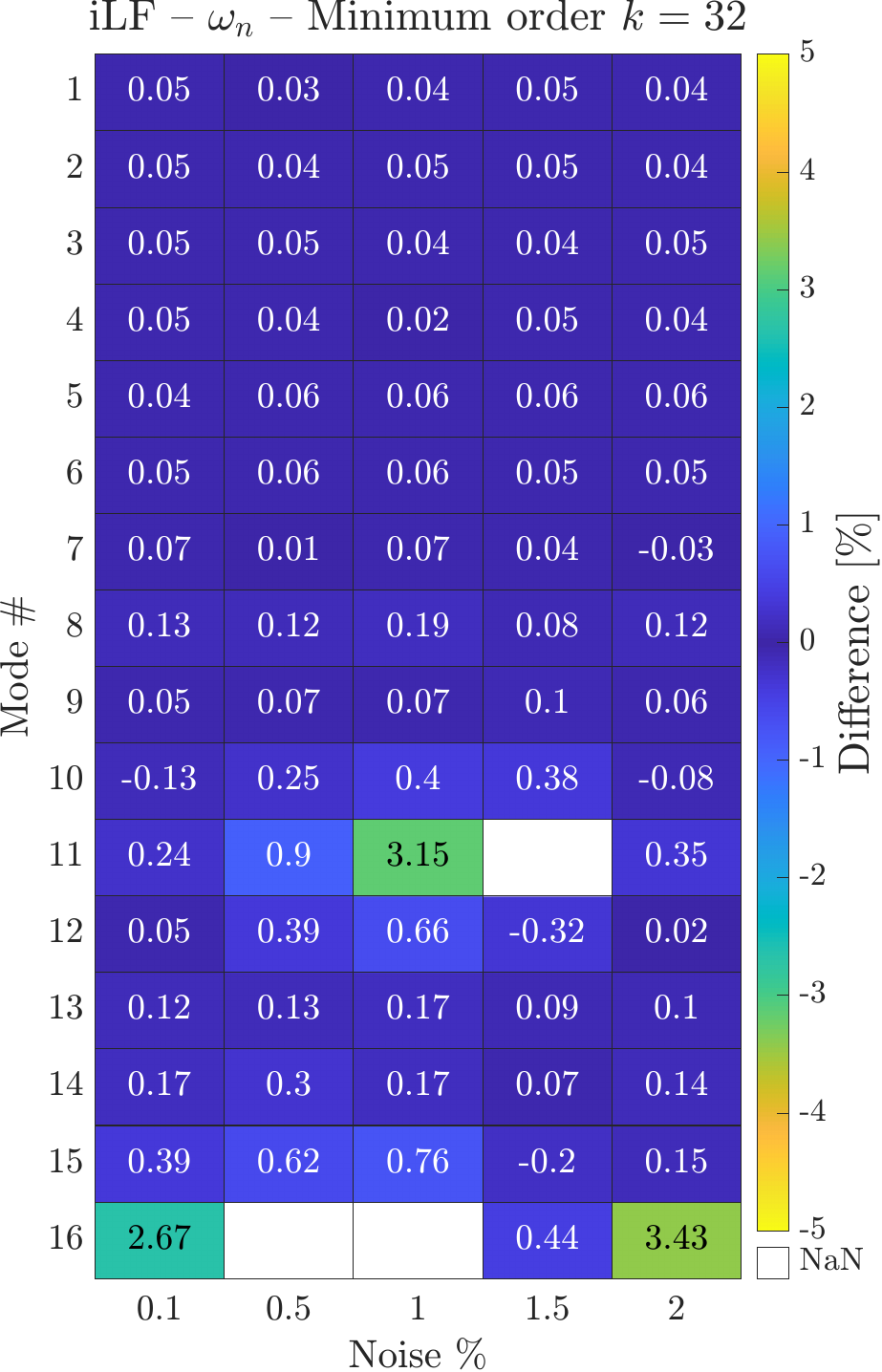}}
		\captionsetup{font={it},justification=centering}
		\subcaption{\label{fig:noise_w}}	
	\end{subfigure}
    \begin{subfigure}[t]{.325\textwidth}
	\centering
		\includegraphics[width=\textwidth,keepaspectratio]{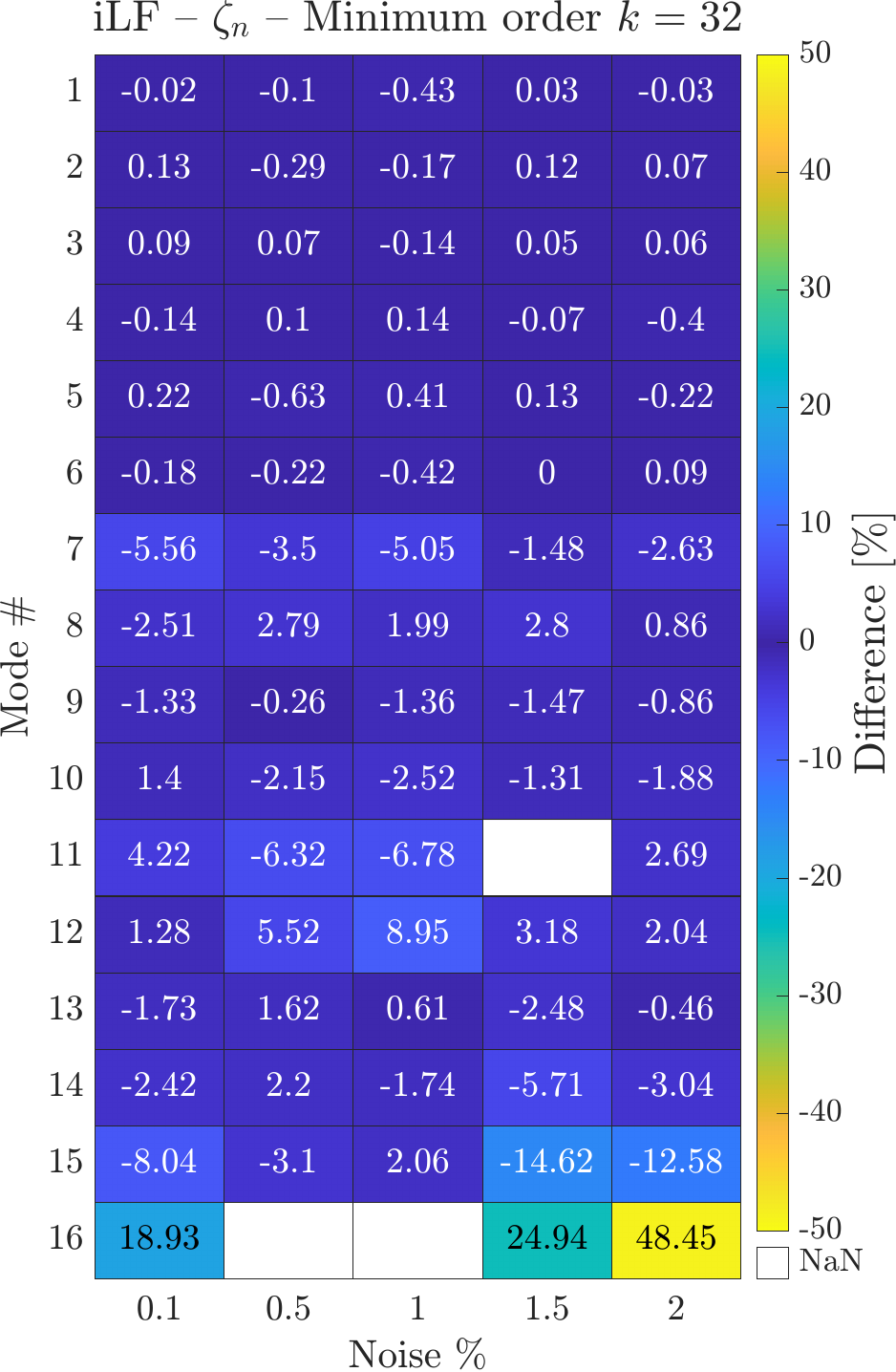}
		\captionsetup{font={it},justification=centering}
		\subcaption{\label{fig:noise_z}}	
	\end{subfigure}
    \begin{subfigure}[t]{.325\textwidth}
	\centering
		{\includegraphics[width=\textwidth,keepaspectratio]{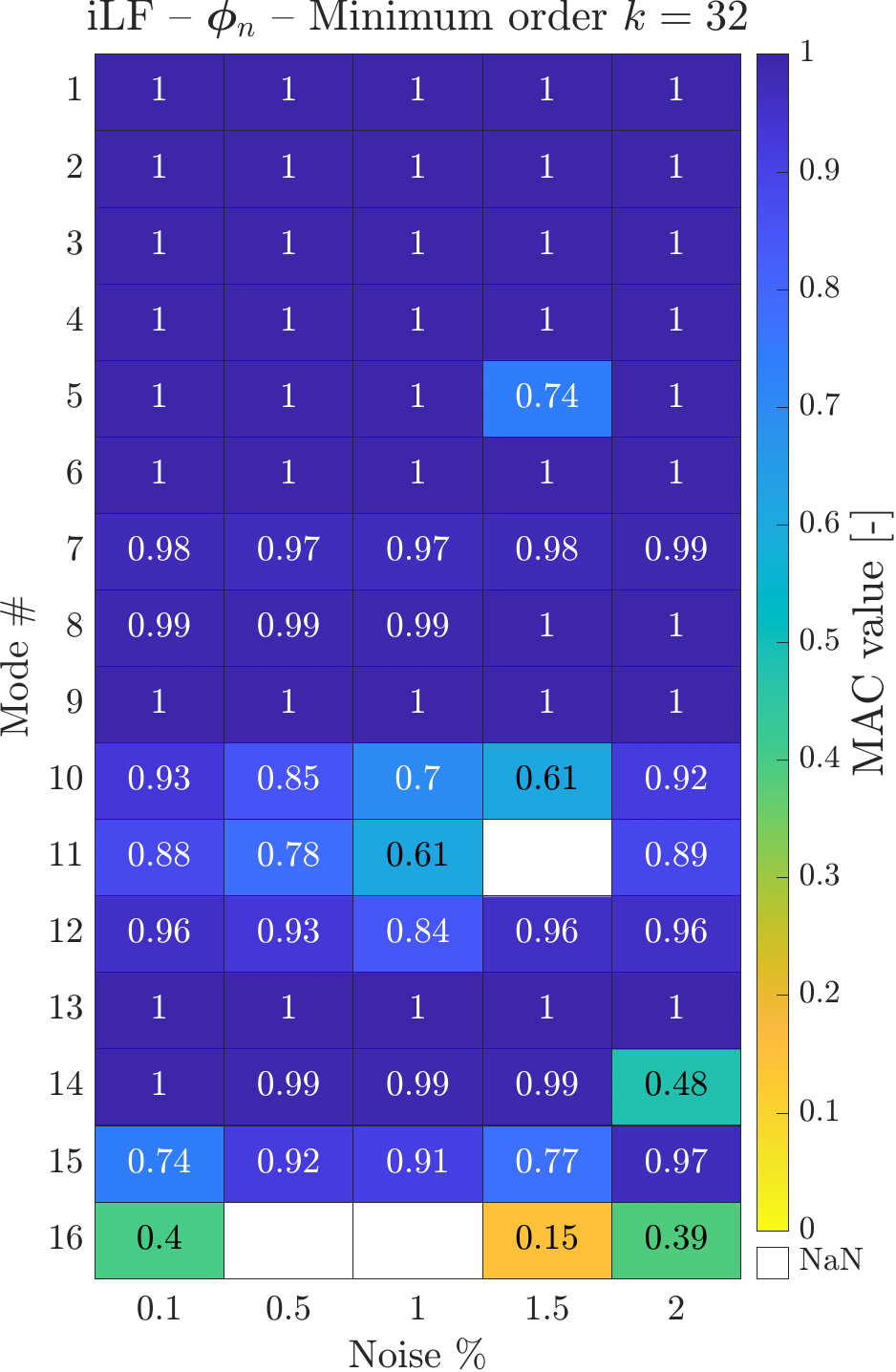}}
		\captionsetup{font={it},justification=centering}
		\subcaption{\label{fig:noise_p}}	
	\end{subfigure}
	\caption{8-element 3D beam: Effects of input-output noise on the identification (difference w.r.t analytical results) of $\omega_n$ (\cref{fig:noise_w}), $\zeta_n$ (\cref{fig:noise_z}), and $\mathbf{\phi}_n$ (\cref{fig:noise_p}) via the iLF at the minimum order $k$.}
	\label{fig:noise}
\end{figure*}

\begin{figure*}[!ht]
\centering
	\begin{subfigure}[t]{.32\textwidth}
	\centering
		{\includegraphics[width=\textwidth,keepaspectratio]{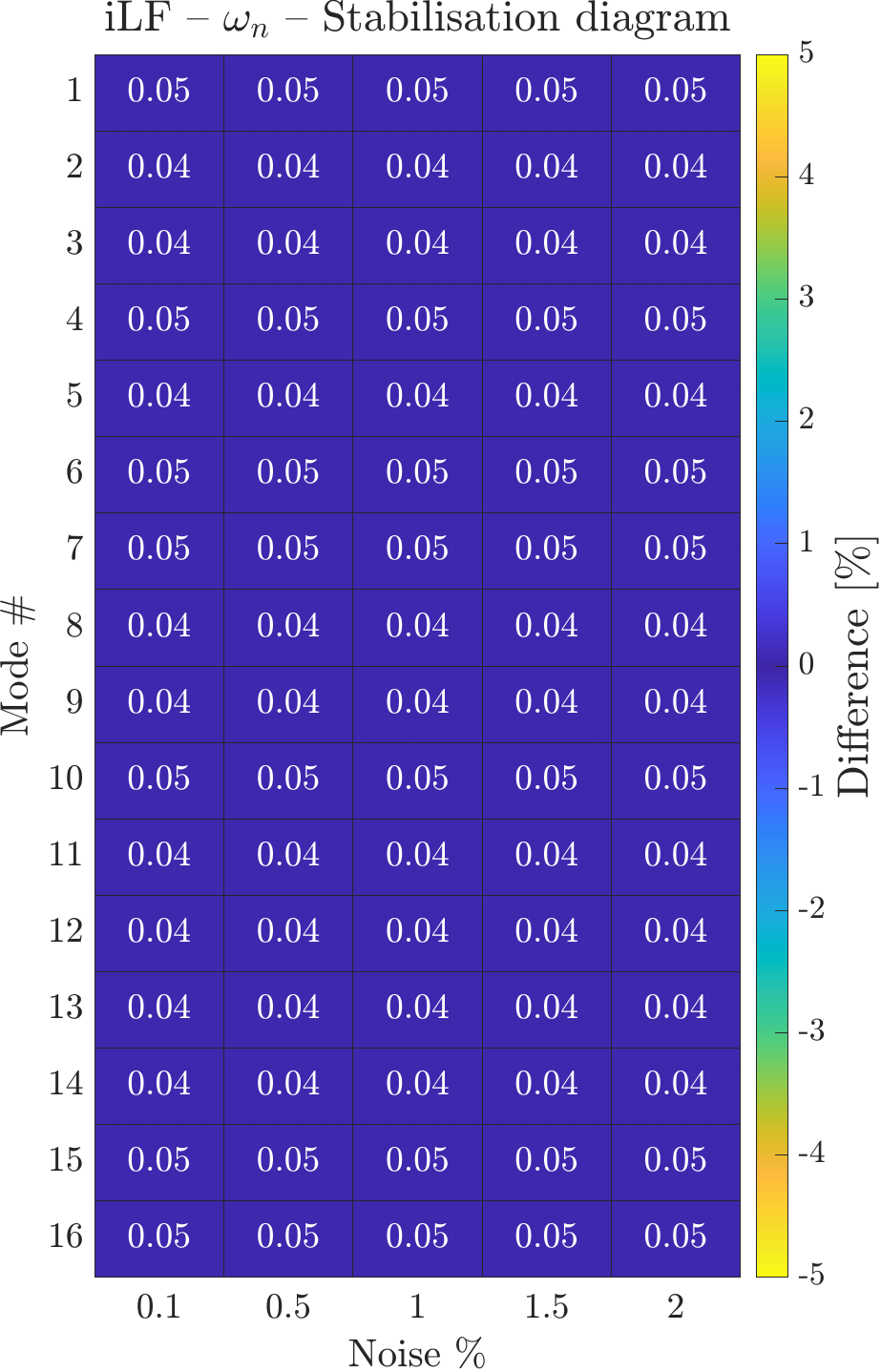}}
		\captionsetup{font={it},justification=centering}
		\subcaption{\label{fig:noise_stab_w}}	
	\end{subfigure}
    \begin{subfigure}[t]{.32\textwidth}
	\centering
		\includegraphics[width=\textwidth,keepaspectratio]{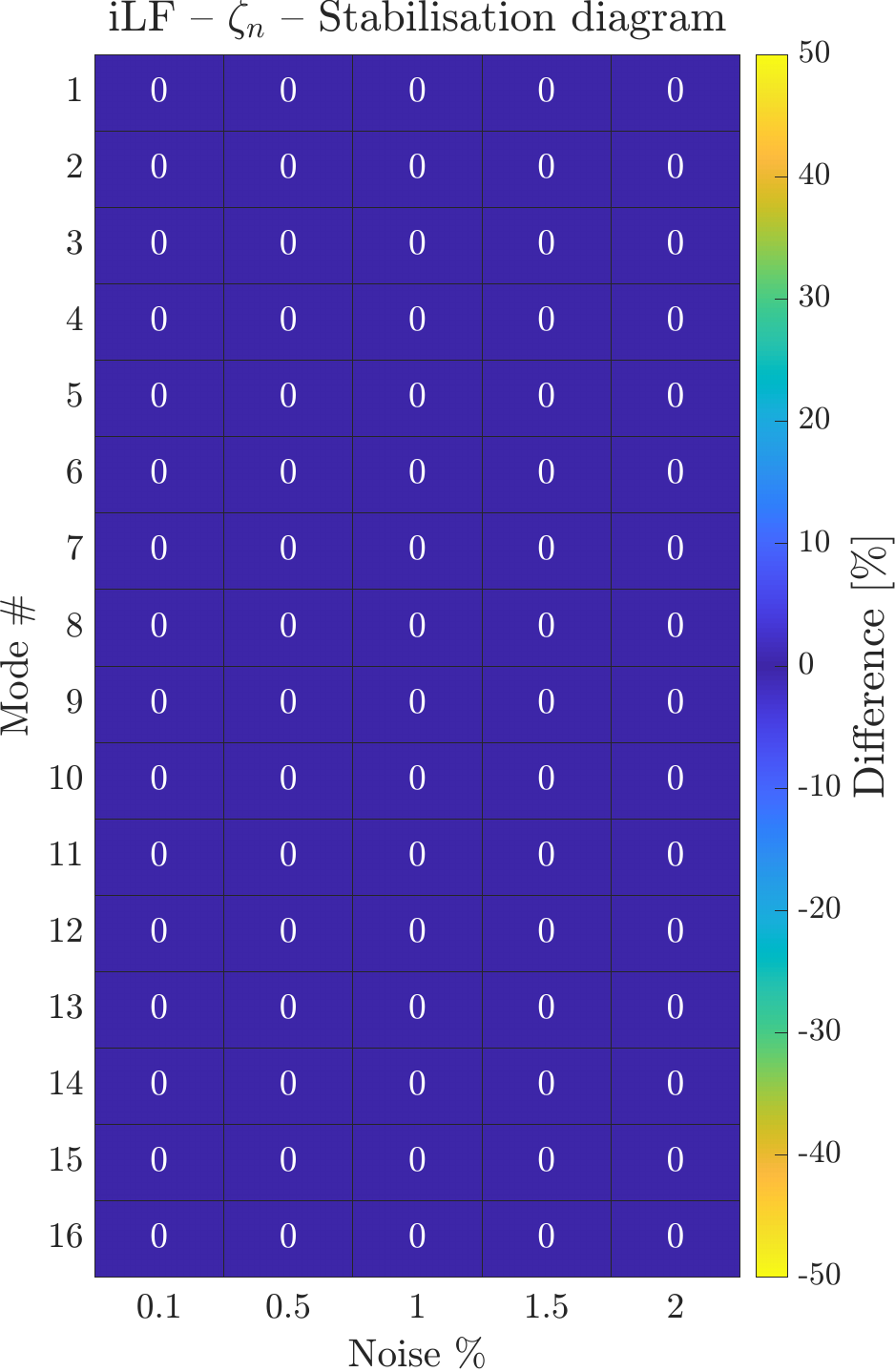}
		\captionsetup{font={it},justification=centering}
		\subcaption{\label{fig:noise_stab_z}}	
	\end{subfigure}
    \begin{subfigure}[t]{.32\textwidth}
	\centering
		{\includegraphics[width=\textwidth,keepaspectratio]{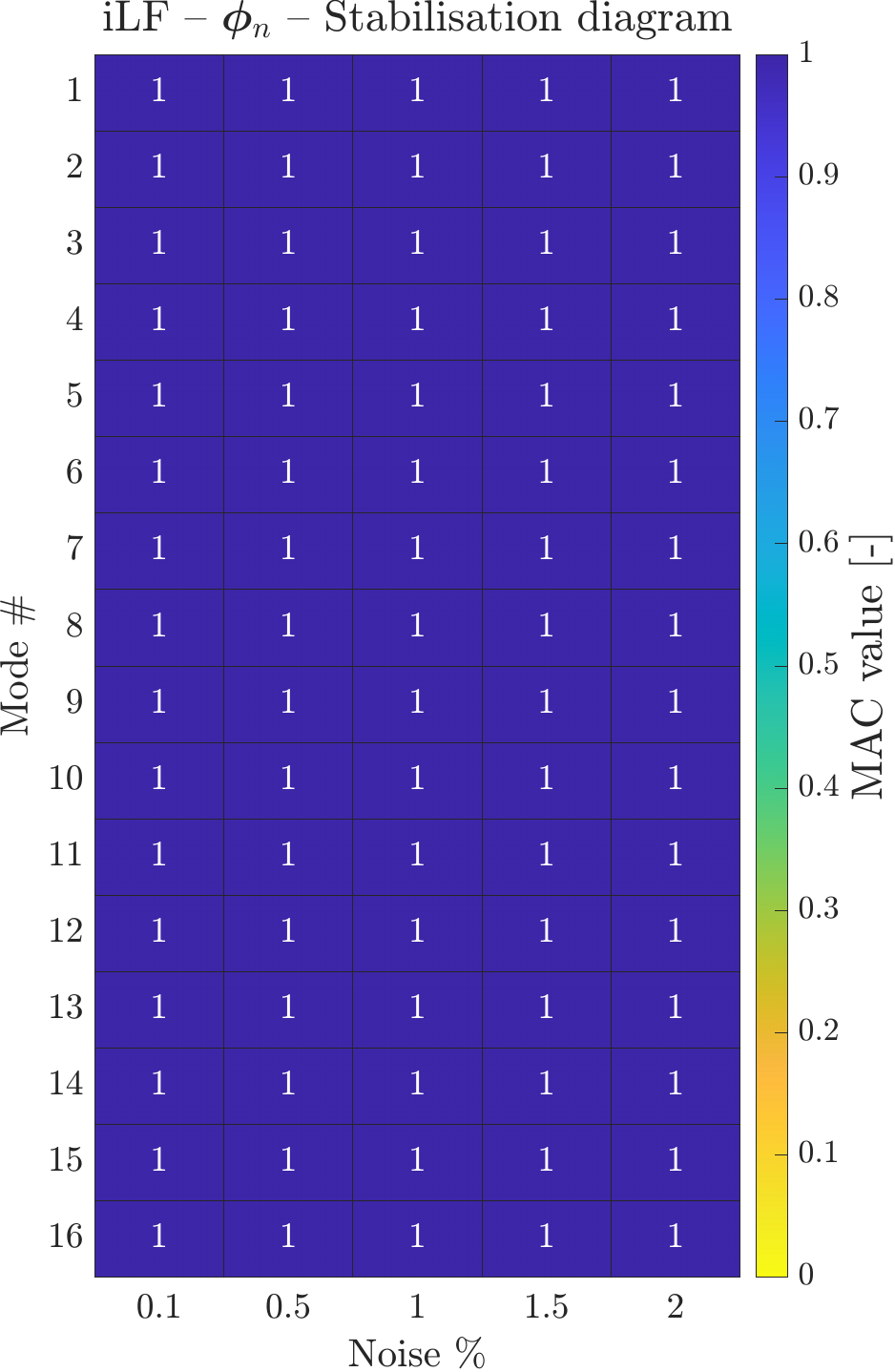}}
		\captionsetup{font={it},justification=centering}
		\subcaption{\label{fig:noise_stab_p}}	
	\end{subfigure}
	\caption{8-element 3D beam: Effects of input-output noise on the identification (difference w.r.t. analytical results) of $\omega_n$ (\cref{fig:noise_stab_w}), $\zeta_n$ (\cref{fig:noise_stab_z}), and $\mathbf{\phi}_n$ (\cref{fig:noise_stab_p}) via the iLF for the stable modes, such that $k$ $\in$ [32,60].\label{fig:noise_stab}}
\end{figure*}

The idea is to compare the modal parameters identified from the noisy measurements to those from the analytical estimations. This comparison is shown in \cref{fig:noise} and \ref{fig:noise_stab}, respectively, for the case of minimum order $k$ and stable modes.

From \cref{fig:noise}, it is clear that the first 15 modes are identified within acceptable error margins. For example, the maximum error in terms of $\omega_n$ is less than 0.5\%, less than 8.5\% for $\zeta_n$ and the lowest MAC value of $\mathbf{\phi}_n$ is 0.74, being 0.8 the minimum for an acceptable correlation. In addition, the first nine modes (exception made for mode \#5 for the 1.5\% noise level) are coherent with the analytically derived values for all noise levels. As expected, the maximum discrepancies are found for the higher modes. This is a common trend for LF-derived techniques, and it was also found in the SIMO application of LF reported in \cite{Dessena2022g}. The most significant issues in modal identification precision are found for the higher modes and the modes close in frequency.

The results of the identification of stable modes from the noised signals are shown in \cref{fig:noise_stab}. Here the results are similar to that of the noiseless case, with minimal variations (less than 0.06\%) in $\omega_n$ and virtually no changes in the identified $\zeta_n$ (differences are all{ -- or close to --} of 0\%) and $\mathbf{\phi}_n$ (all MAC values are 1).
In conclusion, the computational improvement of the iLF with respect to the conventional LF does not hamper its precision and robustness to noise. In particular, it offers the same precision as industry-standard methods, like N4SID, on the numerical cases analysed. In addition, it can identify systems {that} standard methods, like LSCE, struggle with - as it was for {this case study}. Further, experimental dataset verification and validation are carried out in the remainder of the article. 

\section{The BAE Systems Hawk T1A aircraft benchmark}\label{sec:exp}

\begin{figure}[!ht]
\centering
	\begin{subfigure}[t]{.49\textwidth}
	\centering
		{\includegraphics[width=\textwidth,keepaspectratio]{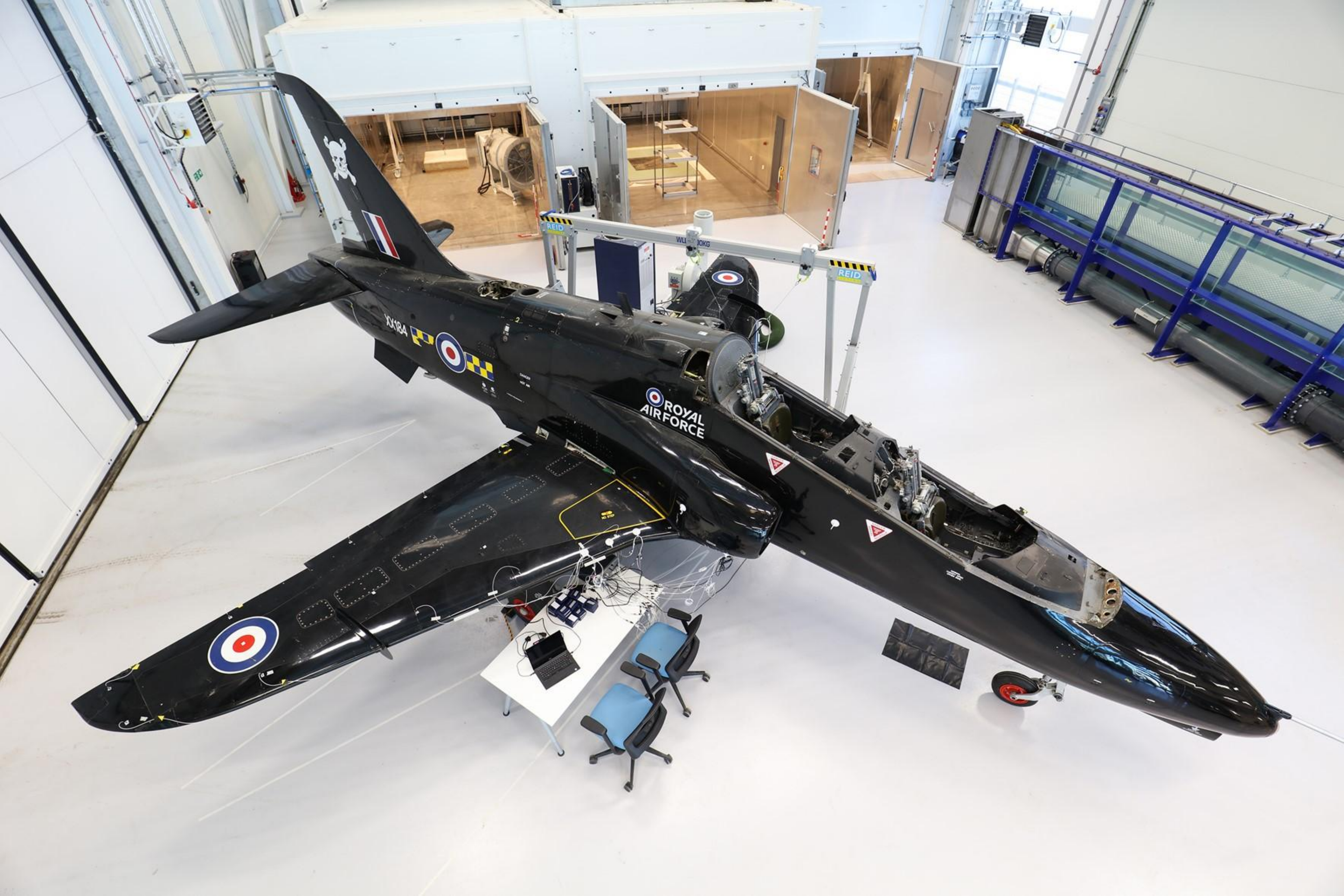}}
		\captionsetup{font={it},justification=centering}
		\subcaption{\label{fig:hawk_a}}	
	\end{subfigure}
    \begin{subfigure}[t]{.49\textwidth}
	\centering
		\includegraphics[width=.88\textwidth,keepaspectratio]{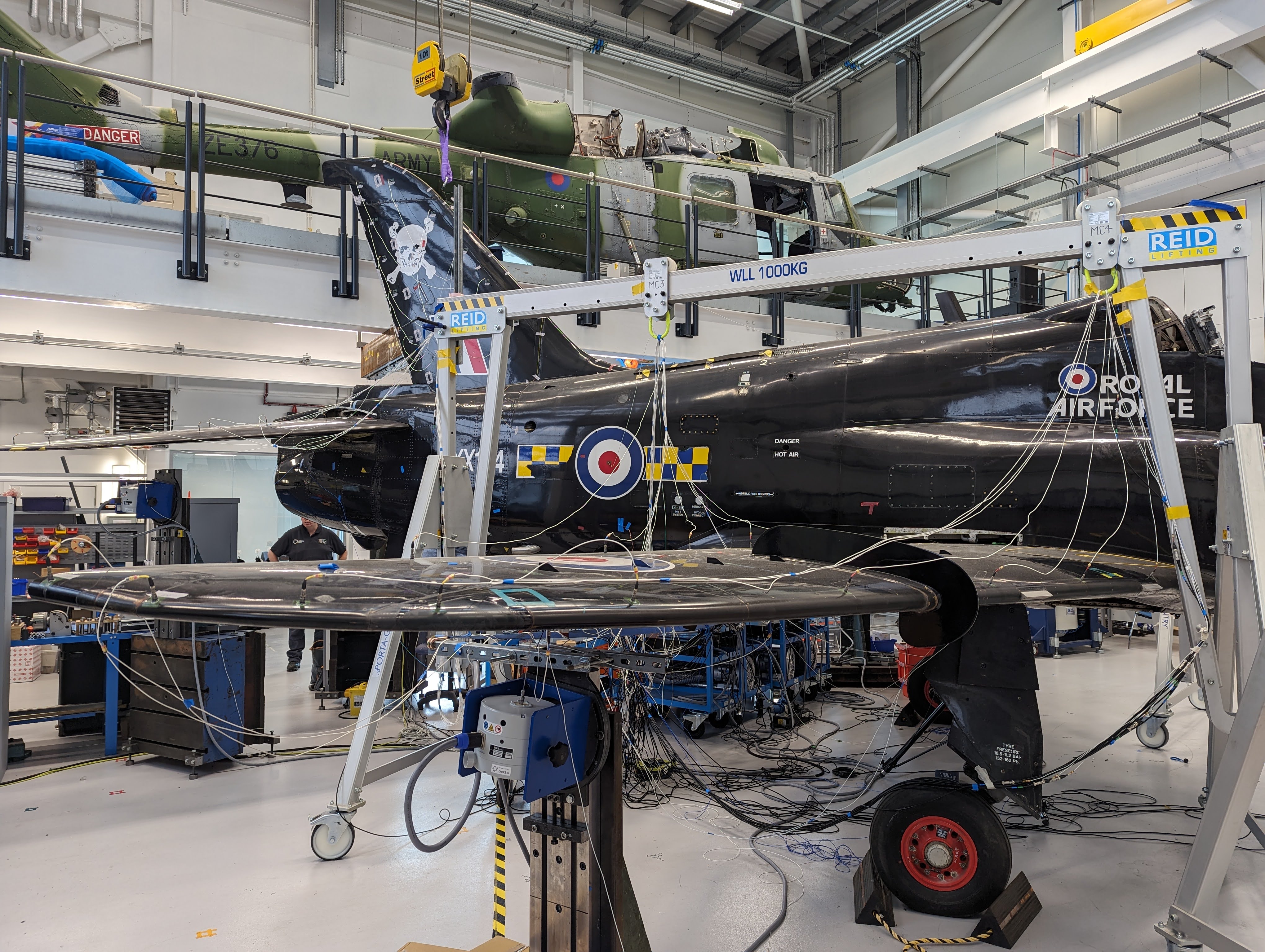}
		\captionsetup{font={it},justification=centering}
		\subcaption{\label{fig:hawk_b}}	
	\end{subfigure}
	\caption{Hawk T1A Aircraft: first (\cref{fig:hawk_a} -- retrieved from \cite{Haywood-Alexander2024}) and second test setup (\cref{fig:hawk_b} -- retrieved from \cite{Wilson2024}).}
	\label{fig:hawk}
\end{figure}

As of the writing of this article, the experimental data from the BAE Systems Hawk T1A aircraft has been recently published by researchers from the Dynamics Research Group at the University of Sheffield. The full aircraft, a Hawk T1A formerly in service with the British Royal Air Force and now decommissioned, is depicted in \cref{fig:hawk}. The craft is shown at the University of Sheffield's Structural Dynamics Laboratory for Verification and Validation (LVV) facilities (Sheffield, UK), where it was tested. Importantly, the data used in this article are firstly described in two separate documents: \cite{Haywood-Alexander2024} and \cite{Wilson2024}. They focus on, respectively, the starboard wing and the whole Hawk airframe structure. Both were used here for experimental validation. These two related datasets serve as a compelling benchmark for dynamic System Identification techniques. 
Nevertheless, as discussed in this Section, some precautions have been considered to ensure the full comparability of data from the two experimental campaigns.

In both cases, only the key aspects of the recorded data used for the algorithm validation will be recalled here. The interested reader can refer to the original sources for further insights.

\subsection{Wing experimental setup and data description}

\begin{figure}[!ht]
\centering
	\begin{subfigure}[c]{.85\textwidth}
	\centering
		{\includegraphics[width=\textwidth,keepaspectratio]{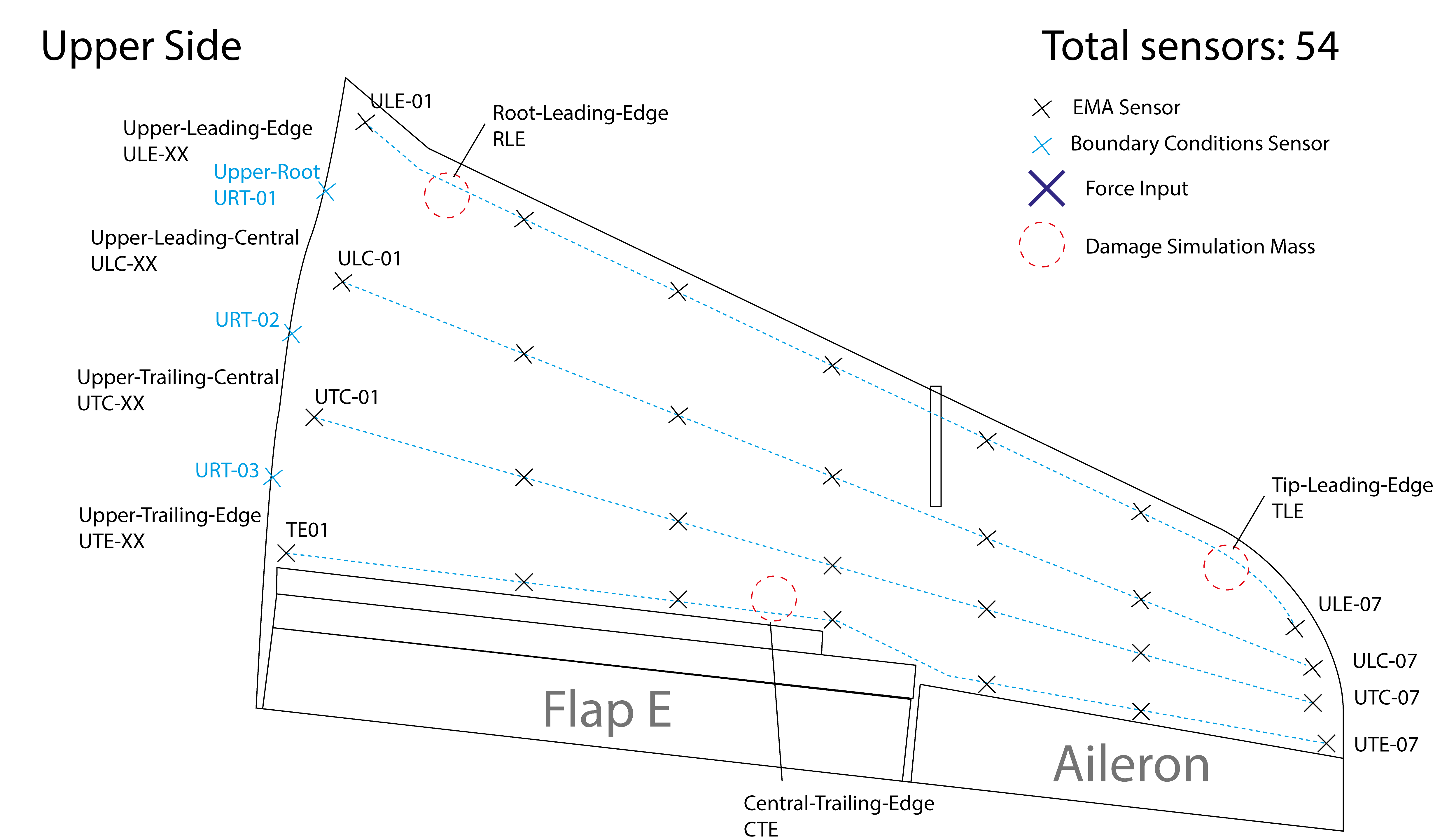}}
		\captionsetup{font={it},justification=centering}
		\subcaption{{\label{fig:aileron_a}}}	
	\end{subfigure}
    \begin{subfigure}[c]{.85\textwidth}
	\centering
		\includegraphics[width=\textwidth,keepaspectratio]{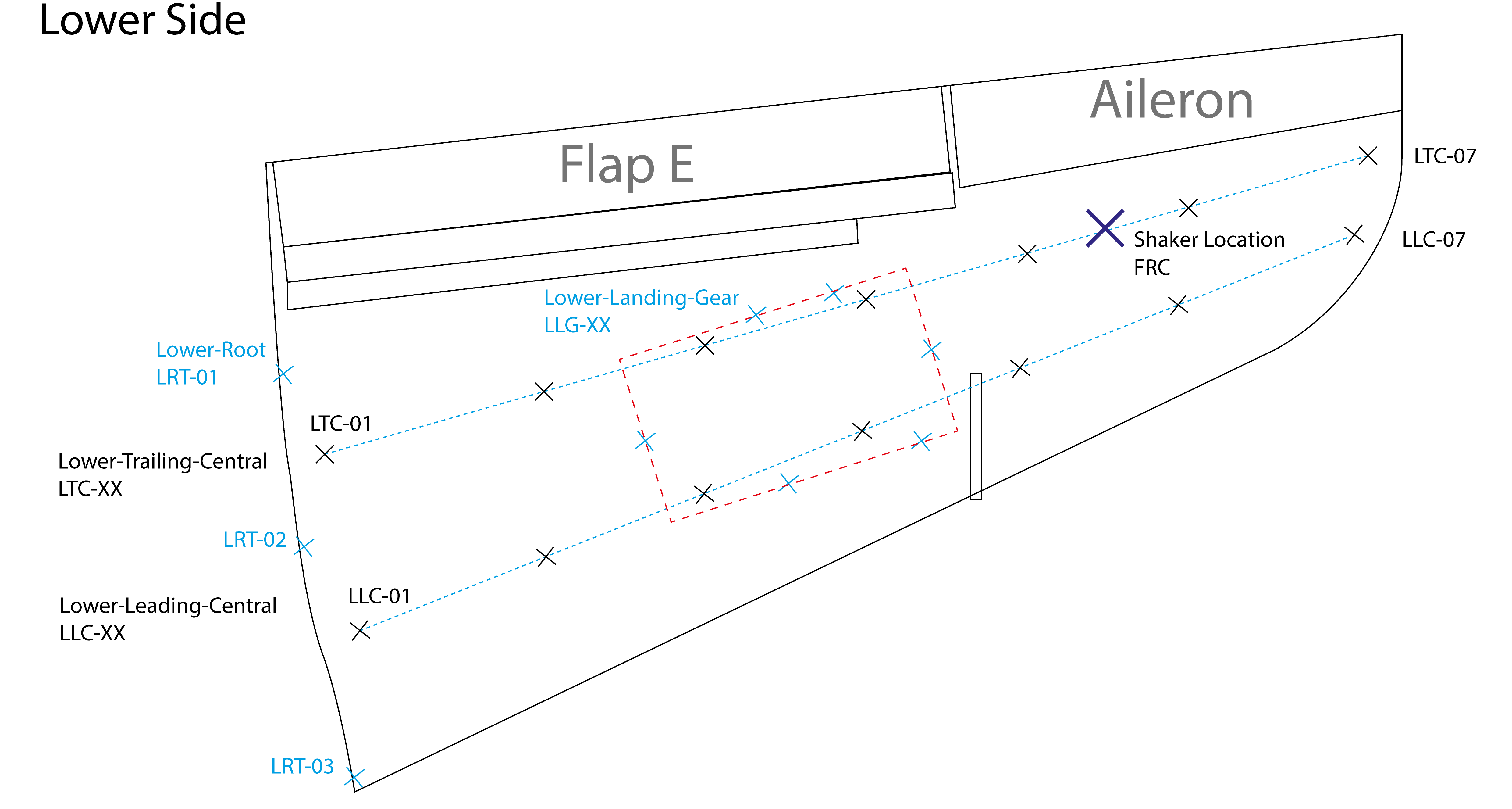}
		\captionsetup{font={it},justification=centering}
		\subcaption{\label{fig:aileron_b}}	
	\end{subfigure}
	\caption{Hawk T1A Aircraft: the complete starboard wing test sensor layout for the first dataset. \Cref{fig:aileron_a} shows the wing upper surface and \cref{fig:aileron_b} the lower surface. Adapted from \cite{Haywood-Alexander2024}).}
	\label{fig:aileron}
\end{figure}

A total of 54 PCB Piezotronics accelerometers (nominal sensitivity: 10  mVg\textsuperscript{-1}) are mounted on the lower and upper surface of the starboard wing, as shown in \cref{fig:aileron}. They are arranged in a grid of three and two lines, respectively, running parallel to the leading edge. Extra sensors are placed at the root boundary condition locations and by the undercarriage door. One Tira TV 51140-MOSP modal shaker, directly attached to a PCB Piezotronics 208C02 force transducer with a sensitivity of 11.11 mVN\textsuperscript{-1}, applies the input driving force on the wing lower surface. Specifically, the acquisitions corresponding to burst random excitation{ – white noise forcing input, with amplitude ramping at the beginning and end of the burst – }with 0.4 V {LMS} input voltage are considered in this study. {A Siemens PLM LMS SCADAS unit was used to generate the input signal, and record the acceleration responses.}

As can be inferred, this dataset only includes one input source. Therefore, it was used to obtain the SISO readings to compare the SISO and SIMO capabilities of the iLF while getting baseline values for system modes.

More details on the wing setup and experiments can be found in \cite{Haywood-Alexander2024}; the data is found in \cite{Haywood-Alexander2023}.

\subsection{Whole airframe setup and data description}

Recalling the structure of the target aircraft under testing conditions, as depicted in \cref{fig:hawk}, the second dataset, available at \cite{Wilson2024DATASET}, is made up of more than 200 test conditions, including different excitation signals and amplitude levels. The same model of modal shakers and force transducers as in the other case were used.

\begin{figure}[!ht]
\centering
	\begin{subfigure}[c]{\textwidth}
	\centering
		{\includegraphics[width=.85\textwidth,keepaspectratio]{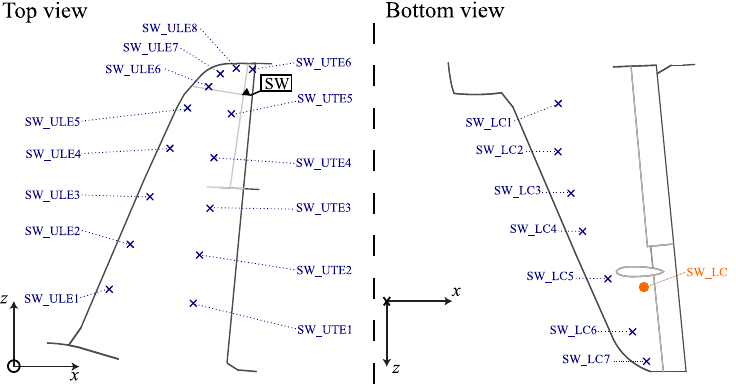}}
		\captionsetup{font={it},justification=centering}
		\subcaption{\label{fig:MC_SENSORS_LOCATIONS_a}}	
	\end{subfigure}
    \begin{subfigure}[c]{\textwidth}
	\centering
		\includegraphics[width=.85\textwidth,keepaspectratio]{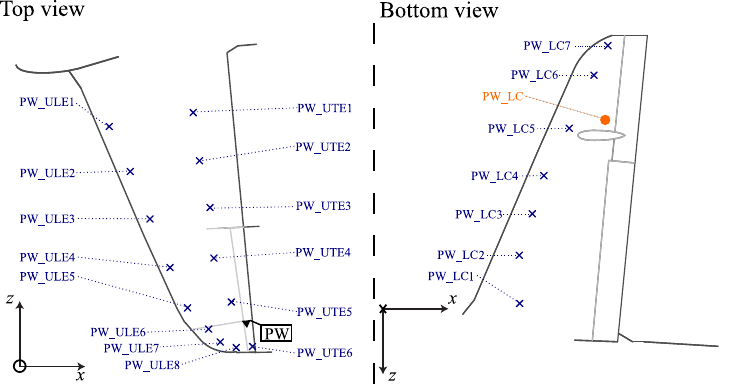}
		\captionsetup{font={it},justification=centering}
		\subcaption{\label{fig:MC_SENSORS_LOCATIONS_b}}	
	\end{subfigure}
	\caption{Hawk T1A Aircraft: the sensor layout for the MIMO tests on the starboard (\cref{fig:MC_SENSORS_LOCATIONS_a}) and port wing (\cref{fig:MC_SENSORS_LOCATIONS_b}). Adapted from \cite{Wilson2024}. Please note: {In the original images of the wings bottom view \cite{Wilson2024}, the sensors were ordered from the wing tip to the root. However, the numbering in the dataset metadata \cite{Wilson2024DATASET} reflects that of the modified figures presented here}.\label{fig:MC_SENSORS_LOCATIONSa}}
\end{figure}

\begin{figure}[!ht]
\centering
    \begin{subfigure}[c]{\textwidth}
	\centering
		{\includegraphics[width=.85\textwidth,keepaspectratio]{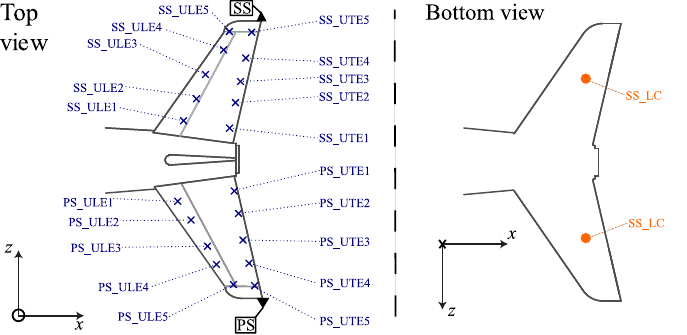}}
		\captionsetup{font={it},justification=centering}
		\subcaption{\label{fig:MC_SENSORS_LOCATIONS_c}}	
	\end{subfigure}
    \begin{subfigure}[c]{\textwidth}
	\centering
		\includegraphics[width=.85\textwidth,keepaspectratio]{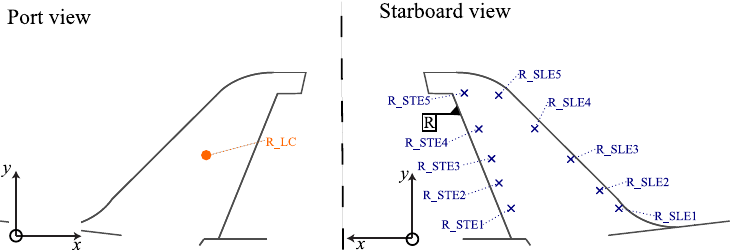}
		\captionsetup{font={it},justification=centering}
		\subcaption{\label{fig:MC_SENSORS_LOCATIONS_d}}	
	\end{subfigure}
	\caption{Hawk T1A Aircraft: the horizontal stabiliser (\cref{fig:MC_SENSORS_LOCATIONS_c}), and the vertical stabiliser (\cref{fig:MC_SENSORS_LOCATIONS_d}). Adapted from \cite{Wilson2024DATASET}.\label{fig:MC_SENSORS_LOCATIONSb}}
\end{figure}

{To ensure comparability with the wing dataset, the recordings corresponding to a WGN input with the same input voltage {(0.4 V - before amplification)} have been used}. {It should be noted, however, that MIMO tests, by definition, result in an overall higher excitation energy. Here, unique drive signals for each of the five shakers were generated by summing sine waves with random phases, followed by iterative adjustments to meet a defined tolerance of 4\% between measured and target power spectral densities. Once aligned, spectral gain masks were stored and used to generate the drive signals, allowing for variable test periods through frequency domain resampling by linear interpolation. In }\cite{Wilson2024},{ no information seems to be explicitly reported on the data collection system and the software used. Nevertheless, these can be inferred to be National Instruments hardware with LabView software, since a similar control algorithm implementation is used in }\cite{Haywood-Alexander2024}{ for the closed-cloop wing tests.}
In this case, five modal shakers are attached to the starboard wing (i.e. the one already discussed above), the port wing, plus the starboard horizontal stabiliser, port horizontal stabiliser, and vertical stabiliser. The exact locations are marked in orange in \cref{fig:MC_SENSORS_LOCATIONSa,fig:MC_SENSORS_LOCATIONSb}. 

\begin{figure}[!ht]
\centering
	\begin{subfigure}[c]{.49\textwidth}
	\centering
		{\includegraphics[width=\textwidth,keepaspectratio]{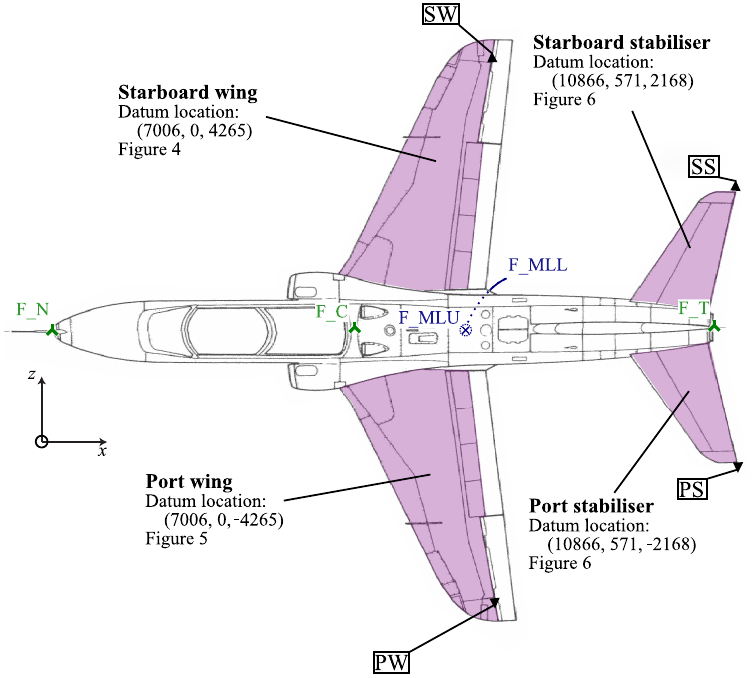}}
		\captionsetup{font={it},justification=centering}
		\subcaption{\label{fig:MC_SENSORS_LOCATIONS2_a}}	
	\end{subfigure}
    \begin{subfigure}[c]{.49\textwidth}
	\centering
		\includegraphics[width=\textwidth,keepaspectratio]{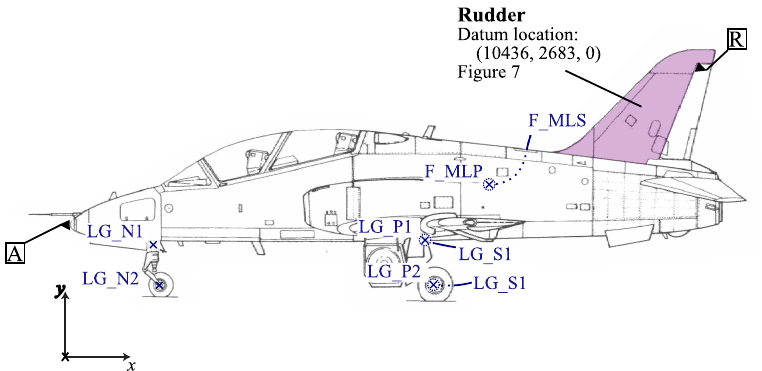}
		\captionsetup{font={it},justification=centering}
		\subcaption{\label{fig:MC_SENSORS_LOCATIONS2_b}}	
	\end{subfigure}
	\caption{Hawk T1A Aircraft: the sensor layout for the MIMO tests on the fuselage and landing gears, seen from above and the port side, \cref{fig:MC_SENSORS_LOCATIONS2_a,fig:MC_SENSORS_LOCATIONS2_b} respectively. Adapted from \cite{Wilson2024DATASET} }
	\label{fig:MC_SENSORS_LOCATIONS2}
\end{figure}

A total of 85 between uni- and tri-axial accelerometers, resulting in 91 output channels, are mounted on the airframe, plus other sensing devices for a total of 139 recording channels, including other-than-accelerometers - namely strain gauges, temperature, and humidity probes, not used in this research work. More details on the accelerometers are found below \cite{Wilson2024DATASET}):

\begin{enumerate}
    \item Both the starboard and the port wing (\cref{fig:MC_SENSORS_LOCATIONS_a,fig:MC_SENSORS_LOCATIONS_b} respectively) have two rows of uniaxial accelerometers on the top side (14 sensors), parallel to the leading and trailing edges, and one on the bottom side (7 sensors), parallel to the leading edge only, thus totalling 21 output channels per wing;
    \item The two sides of the horizontal stabiliser are populated with two rows of five uniaxial sensors at both the rear and front side, totalling 20 per side (located exclusively on the upper; see \cref{fig:MC_SENSORS_LOCATIONS_c});
    \item Similarly, the vertical stabiliser sensors are placed only on the starboard side, again with 10 sensors equally distributed along the leading and trailing edge (\cref{fig:MC_SENSORS_LOCATIONS_d});
    \item All the accelerometers on the wings and stabilisers have the same nominal sensitivity of 10 mVg\textsuperscript{-1}.
    \item  Three triaxial PCB Piezotronics accelerometers with higher nominal sensitivity (100 mVg\textsuperscript{-1}) are placed at the nose, cockpit, and tail of the fuselage (indicated in green in Figure \cref{fig:MC_SENSORS_LOCATIONS2_a})
    \item Another set of four high-sensitivity (100 mVg\textsuperscript{-1}) uniaxial PCB Piezotronics accelerometers are placed on the fuselage midpoint on the upper, lower, port, and starboard sides (F\_MLU, F\_MLL, F\_MLP, and F\_MLS in Figure \cref{fig:MC_SENSORS_LOCATIONS2_a});
    \item Finally, two 100 mVg\textsuperscript{-1} uniaxial sensors are placed on each landing gear wheel axle and above the hydraulic suspension, totalling six.
 \end{enumerate}
    
The 86th sensor, a triaxial accelerometer laid on the ground as a reference, is obviously not used here for SI. Apart from this, all the other output channels are selected for MIMO System Identification, in conjunction with the five input channels. Conversely, the fibre-Bragg grating strain gauges - described in \cite{Wilson2024DATASET} - are not considered here. 

More details on the full aircraft setup and experiments can be found in \cite{Wilson2024}.

\subsubsection{Pre-processed FRFs}

Regarding the processed data, the Frequency Response Functions (FRFs) are used. These are retrieved for the first dataset, where they are already available in the frequency domain, and computed for the second dataset. 

It is important to remark that for the FRFs for the SISO case (first dataset), the analytical definition was used by the authors of the database, following \eqref{eq:9}, thus using  $Y(\omega)$ i.e. the Fourier Transform of the output and $X(\omega)$ i.e. the Fourier Transform of the input. This is also consistent with the procedure performed here in this paper for the preliminary numerical analyses (which will be detailed later).

Conversely, to ensure more noise robustness, the H1 estimators are used in lieu of the actual FRF formulation for the second dataset (whole airframe). These are defined as 
$$H_1(\omega)=S_{YX}(\omega)/S_{XX}(\omega)$$
i.e. the ratio between the cross-spectral density of the output with the input and the auto-spectral density of the input. 
More specifically, in the dataset used here, the cross- and auto-spectral densities are computed via a short-time Fourier transform (STFT) with a Hann window, a segment length of 16384 and an overlap of 8192. This was mentioned in \cite{Wilson2024} and replicated in this work.

$H_1$ estimators are indeed appreciated for their capability to reduce output measurement noise, which more significantly affects the valleys of FRFs with signal leakage. However, it is well-known that, while the FRF estimators allow for averaging over multiple acquisitions and provide more accurate estimates in absolute value, this is achieved at the cost of losing the phase information {in the auto-spectra. Nevertheless, this information is retained in the cross-spectra} (see e.g. \cite{Avitabile2017} Chapter 3.13). 

Note that data for 10 repetitions of the same test is available for each test in both datasets. Hence, as suggested by the authors of the datasets \cite{Champneys2023}, the mean of the FRFs over each repetition is considered in all computations to improve the signal-to-noise ratio. 

\subsubsection{Further considerations}

Before continuing, there are two key aspects worth mentioning. 
The first one is that the sensor layout for the starboard wing used in \cite{Wilson2024DATASET} is similar but not identical to the one of \cite{Haywood-Alexander2023}, as e.g. only two lines of sensors were used on the upper side (one on the lower side). As it is better discussed in the \cref{sec:exp}, the nearest output channels were selected to compare SISO, MISO, and MIMO tests between these two datasets.

The second aspect is that for the experiments on the whole airframe with WGN input, the excitation bandwidth was 5-256 Hz
and the sampling rate was 2048 Hz. On the other hand, the first experimental campaign used a sampling frequency of 512 Hz, but it was bandlimited between {5-256 Hz}. Therefore, all analyses have been here limited to $f_{up} = 256$ Hz.

Before moving forward, it is necessary to say that LSCE will not be considered for the next study, as it did not show to be able to cope with close-in-frequency modes well.

\subsection{Modal parameters estimation}
\begin{figure}[!ht]
\centering
	\begin{subfigure}[t]{.65\textwidth}
	\centering
		{\includegraphics[width=\textwidth,keepaspectratio]{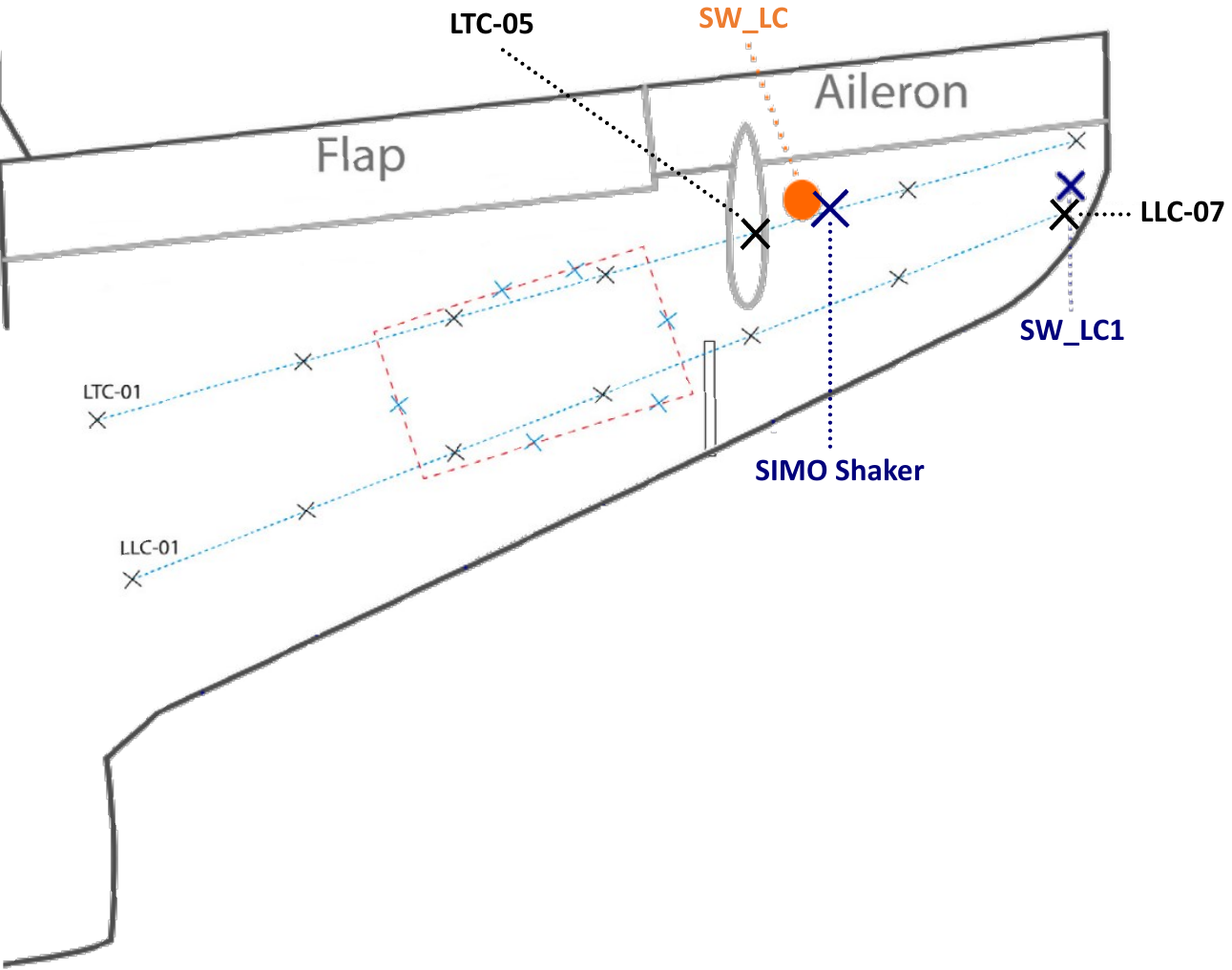}}
		\captionsetup{font={it},justification=centering}
		\subcaption{\label{fig:si_wing}}	
	\end{subfigure}
    \begin{subfigure}[t]{.65\textwidth}
	\centering
		\includegraphics[width=\textwidth,keepaspectratio]{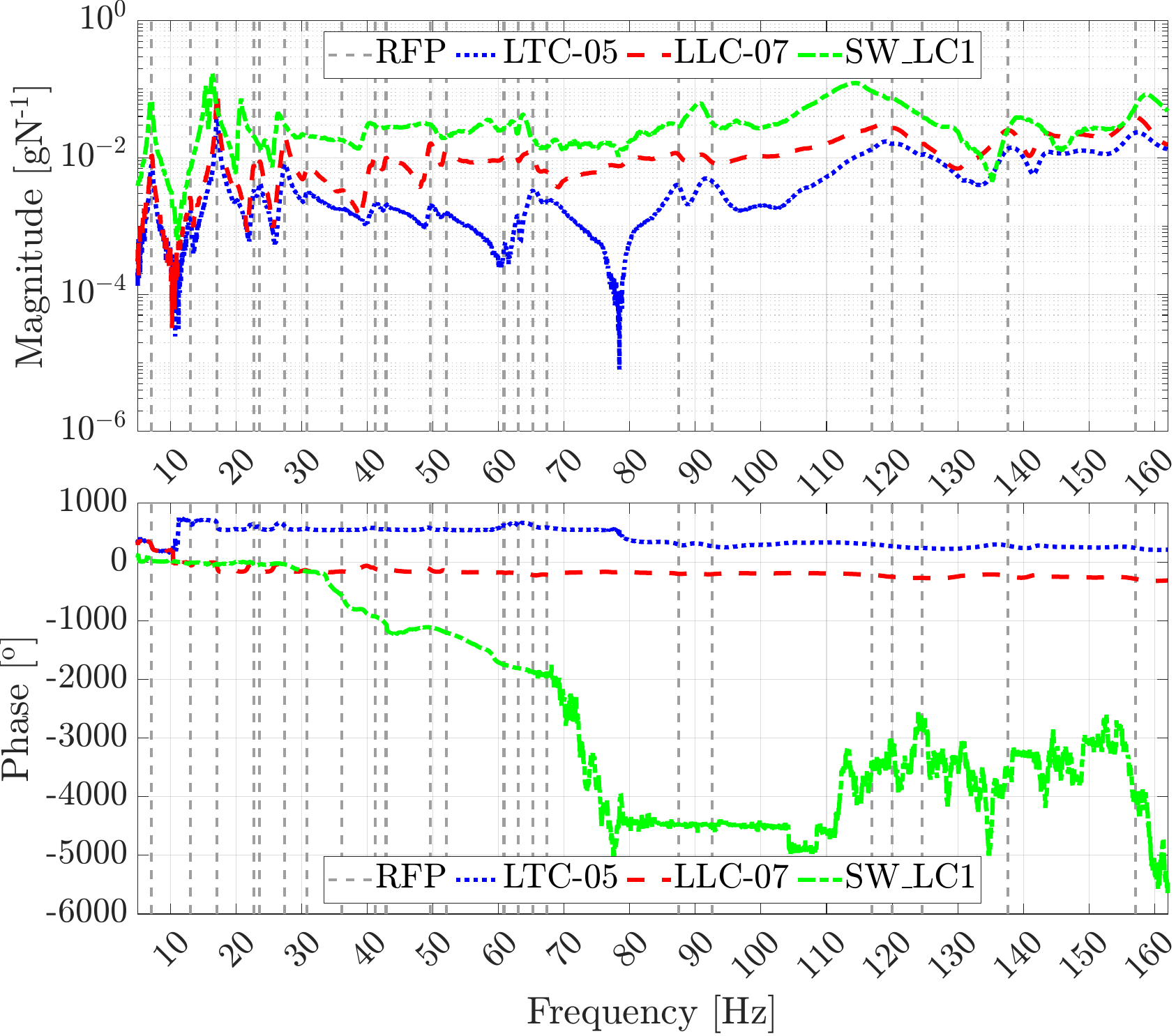}
		\captionsetup{font={it},justification=centering}
		\subcaption{\label{fig:frf_wing}}	
	\end{subfigure}
	\caption{Hawk T1A Aircraft: \cref{fig:si_wing} shows the superimposition of the sensor and shaker locations of the tests in \cite{Haywood-Alexander2024,Wilson2024}, while \cref{fig:frf_wing} show the FRFs of selected channels from the data in \cite{Haywood-Alexander2024,Wilson2024} superimposed to the SISO result in \cite{Haywood-Alexander2024}.}
	\label{fig:3_stab}
\end{figure}

Previous MIMO modal testing work on the aircraft does not explicitly report any results to be used as an external benchmark \cite{Wilson2024}. Nevertheless, data is available from the SIMO modal testing of the aircraft starboard wing \cite{Haywood-Alexander2024}. However, the results presented here are only for the SISO identification via RFP from the FRF for channel LTC-05 (\cref{fig:si_wing}), with no further numerical values included. It is clear that, for getting an idea of where (in frequency) the modes of the full aircraft are, this approach is limited by the data considered. In addition, it should be noted that the second to last mode (mode \#22 according to \cite{Haywood-Alexander2024}) is very likely a repetition of the previous mode (potentially a typo).  Nevertheless, it is possible to retrieve its actual value, as the authors shared the code used to obtain those results in \cite{Champneys2023}. 

In \cite{Haywood-Alexander2024}, ranges for the natural frequencies of the starboard wing are supplied to be used to identify the modes via RFP; hence, the identification is carried out in a peak-picking fashion rather than using higher-order models. Thus, the idea is to compare this SISO result with a MISO result from the same, or close, sensor. To do this, the sensors position graphs of the starboard wing supplied in \cite{Haywood-Alexander2024} (SIMO) and \cite{Wilson2024} (MIMO) are superimposed in \cref{fig:si_wing}. There, it is shown that the shaker is located in (almost) the same position in the two tests; however, the sensor row of LTC-05 is not present in the MIMO sensor layout. Hence, another sensor, with a counterpart in the MIMO test, is identified. In the SIMO test, this sensor is known as LLC-07, while in the MIMO test, {it is identified} as SW\_LC1. 

In \cref{fig:frf_wing}, the FRFs of these three sensors and the lines relative to the RFP identification shown in \cite{Haywood-Alexander2024} are presented. Just from a simple inspection by a trained eye, one can see that more modes possibly exist outside the given frequency bands. The idea is to find these modes by carrying out a SIMO modal analysis for the LTC-05 and LLC-07 channels and an MISO modal analysis on the SW\_LC1 channel before doing the full (91 channels) MIMO modal analysis.

With this in mind, the idea is to do a SISO modal analysis on the frequency bands in \cite{Haywood-Alexander2023} with N4SID and iLF. The frequency bands considered are the following:\\
$\omega_n$ $\in$ [5, 9]; [12, 14]; [15, 19]; [22, 24]; [26, 30]; [30.5, 31]; [35, 37]; [40.5, 44]; [48.5, 54]; [60, 62]; [62.5, 63.5]; [63.8 70]; [86, 90]; [92, 100]; [112, 118.5]; [119.5, 122]; [122, 125]; [135, 138]; [154, 162].

The expectation is that N4SID and iLF will not perform as well as RFP in this task as they are more holistic methods implemented for more typical scenarios with multiple outputs and inputs rather than being used in a peak-picking fashion. Then, the SIMO and MISO analyses outlined above follow.

The identified  $\omega_n$ and $\zeta_n$, for the different cases and sensors, are shown, respectively, in \cref{tab:wing_freq,tab:wing_damp}. The results marked with (*) are those erroneously reported in \cite{Haywood-Alexander2024} and retrieved with the authors-provided \cite{Champneys2023} RFP implementation. Please note that only frequencies 5-160 Hz are inspected in all cases for coherence with \cite{Haywood-Alexander2024}.

\begin{table*}[!hbt]
\centering
\caption{Natural frequencies identified by iLF, N4SID and RFP alongside the results in \cite{Haywood-Alexander2024} for the channels specified in \cref{fig:si_wing}. The values marked with (*) are those mistakenly reported in \cite{Haywood-Alexander2024} and corrected with the code provided in \cite{Champneys2023}.\label{tab:wing_freq}}
\resizebox{.75\textwidth}{!}{  
\begin{tabular}{lc c cc c cc cc cc}
    \hline
    \multicolumn{12}{c}{\textbf{Natural Frequency} [Hz]}\\\hline
    \textbf{Case} & \multicolumn{6}{c}{{SISO}} & \multicolumn{2}{c}{{SIMO}} & \multicolumn{2}{c}{{MISO}} & {MIMO} \\\cline{2-7}\cline{10-11}
    \textbf{Sensor}(s) & \multicolumn{3}{c}{{LTC-05}} & \multicolumn{3}{c}{{LLC-07}} & \multicolumn{2}{c}{{LTC-05 \& LLC-07}} & \multicolumn{2}{c}{{SW\_LC1}} & {All} \\\cline{2-4}\cline{8-9}\cline{12-12}
    \textbf{Mode} & {RFP} \cite{Haywood-Alexander2024} & {N4SID} & {iLF} & {RFP} & {N4SID} & {iLF} & {N4SID} & {iLF} & {N4SID} & {iLF} & {iLF} \\
    \hline
    \emph{1} & 7.09 & 7.10 & 7.11 & 7.11 & 7.10 & 7.11 & - & 7.13 & - & 6.97 & 6.98 \\
    \emph{2} & 13.05 & 13.02 & 13.08 & 13.06 & 13.02 & 13.08 & - & 13.13 & - & 15.47 & 15.43 \\
    \emph{3} & 17.10 & 17.10 & 17.10 & 17.10 & 17.10 & 17.10 & - & 17.11 & - & 16.30 & 16.32 \\
    \emph{4} & - & - & - & - & - & - & - & - & - & - & 17.42 \\
    \emph{5} & - & - & - & - & - & - & - & 19.93 & - & 20.60 & 20.68 \\
    \emph{6} & 22.66 & 22.59 & 22.86 & 22.65 & 22.59 & 22.86 & - & 22.65 & - & - & 22.16 \\
    \emph{7} & - & - & - & - & - & - & - & - & - & - & 25.36 \\
    \emph{8} & 23.56 & 23.58 & 23.67 & 23.53 & 23.58 & 23.67 & - & 23.57 & - & 24.54 & 24.20 \\
    \emph{9} & 27.35 & 27.33 & 27.33 & 27.35 & 27.33 & 27.33 & - & 27.35 & - & 26.24 & 26.29 \\
    \emph{10} & 30.77 & 30.39 & 30.66 & 30.66 & 30.39 & 30.66 & - & 30.76 & - & - & - \\
    \emph{11} & 36.06 & 36.04 & 36.77 & 36.06 & 36.04 & 36.77 & - & - & - & - & - \\
    \emph{12} & 41.16 & 41.24 & 42.88 & 40.99 & 41.24 & 42.88 & - & 41.22 & - & 40.36 & 40.18 \\
    \emph{13} & 42.85 & 42.82 & 43.54 & 42.78 & 42.82 & 43.54 & - & 42.86 & - & - & 43.29 \\
    \emph{14} & - & - & - & - & - & - & - & - & - & - & 46.90 \\
    \emph{15} & 49.61 & 49.62 & 49.78 & 49.47 & 49.62 & 49.78 & - & 49.48 & - & - & 50.00 \\
    \emph{16} & 52.07 & 52.37 & 51.87 & 51.67 & 52.37 & 51.87 & - & 51.74 & - & 53.22 & 53.38 \\
    \emph{17} & 60.81 & 61.46 & 61.59 & 61.92 & 61.46 & 61.59 & - & 60.72 & - & 59.36 & - \\
    \emph{18} & 63.01 & 62.74 & 62.88 & 62.96 & 62.74 & 62.88 & - & 62.61 & - & - & - \\
    \emph{19} & 65.23 & 65.21 & 65.05 & 65.22 & 65.21 & 65.05 & - & 64.97 & - & 63.52 & - \\
    \emph{20} & 67.39 & 68.08 & 66.39 & 67.66 & 68.08 & 66.39 & - & 67.91 & - & - & 67.59 \\
    \emph{21} & - & - & - & - & - & - & - & - & - & 68.48 & 68.75 \\
    \emph{22} & - & - & - & - & - & - & - & - & - & 72.23 & 72.17 \\
    \emph{23} & - & - & - & - & - & - & - & - & - & 79.56 & - \\
    \emph{24} & - & - & - & - & - & - & - & - & - & 81.64 & 81.02 \\
    \emph{25} & - & - & - & - & - & - & - & - & - & 83.88 & 83.91 \\
    \emph{26} & 87.41 & 87.33 & 87.20 & 87.47 & 87.33 & 87.20 & - & 87.14 & - & 85.35 & - \\
    \emph{27} & 92.54 & 90.05 & 92.64 & 91.33 & 90.05 & 92.64 & - & 91.59 & - & 90.75 & - \\
    \emph{28} & - & - & - & - & - & - & - & - & - & 94.11 & 94.70 \\
    \emph{29} & - & - & - & - & - & - & - & - & - & 98.25 & 97.46 \\
    \emph{30} & - & - & - & - & - & - & - & 100.18 & - & - & - \\
    \emph{31} & - & - & - & - & - & - & - & - & - & - & 102.15 \\
    \emph{32} & - & - & - & - & - & - & - & - & - & - & 106.54 \\
    \emph{33} & 116.90 & 116.39 & 117.71 & 117.11 & 116.39 & 117.71 & - & 114.33 & - & 115.33 & 115.05 \\
    \emph{34} & 120.00 & 120.37 & 121.45 & 119.93 & 120.37 & 121.45 & - & 119.73 & - & - & 120.95 \\
    \emph{35} & 124.60 & 120.72 & 123.73 & 124.91 & 120.72 & 123.73 & - & 124.62 & - & - & 125.38 \\
    \emph{36} & - & - & - & - & - & - & - & - & - & 128.68 & 128.98 \\
    \emph{37} & - & - & - & - & - & - & - & - & - & 132.21 & - \\
    \emph{38} & 137.66 (*) & 138.83 & 137.38 & 137.62 & 138.83 & 137.38 & - & 137.91 & - & 139.50 & 137.58 \\
    \emph{39} & - & - & - & - & - & - & - & 142.55 & - & - & - \\
    \emph{40} & - & - & - & - & - & - & - & 149.35 & - & 147.83 & 149.67 \\
    \emph{41} & - & - & - & - & - & - & - & - & - & - & 151.89 \\
    \emph{42} & 157.10 & 156.06 & 156.79 & 157.14 & 156.21 & 157.70 & - & 157.07 & 158.77 & 158.38 & 156.92 \\
    \hline
\end{tabular}
}
\end{table*}
The results presented in \cite{Haywood-Alexander2024} for the band-limited SISO identification via RFP of the LTC-05 sensor of the starboard wing dataset show that 23 modes are found for the wing. Of course, these modes are representative of a local state, and more modes are found when the full aircraft is tested. After all the SISO, SIMO, MISO, and MIMO identifications, {a total of} 42 modes are identified in at least one of the cases. Hence, from now on, these modes are referred to with the numbering shown in \cref{tab:wing_freq,tab:wing_damp}. 

In terms of $\omega_n$ for the SISO identifications, N4SID and iLF always match the frequencies identified by RFP in LTC-05 and LLC-07, except for modes \# 20, 27, and 35, where N4SID underestimates $\omega_n$. The band-limited SISO approach works well when dealing with known frequency bounds, but this is never almost the case in experimental dynamics; thus, higher-order models and means of discerning between numerically spurious modes and real modes are needed. In the SIMO, MISO, and MIMO tests, this is tackled with stabilisation diagrams \cite{Peeters2004}, where subsequent (in terms of ascending model order $k$) identified $\omega_n$, $\zeta_n$, and $\mathbf{\phi}_n$ are checked for stability to verify their physical meaningfulness. The SIMO identification obtained by iLF can identify 4 more modes (\# 5, 30, 39, and 40) but lose out on mode \# 11. These are expected because as one gets a more complete distribution of data along the system, more dominant (in terms of energy amplitude) modes can overshadow weaker modes (usually local or componential) close in frequency. However, a more holistic approach allows to identify modes which are not as apparent in other parts of the structure. This is a recurrent theme as more modes are identified as the test case involves more outputs and channels. 

\begin{figure*}[!ht]
\centering
	\begin{subfigure}[t]{.495\textwidth}
	\centering
		{\includegraphics[width=\textwidth,keepaspectratio]{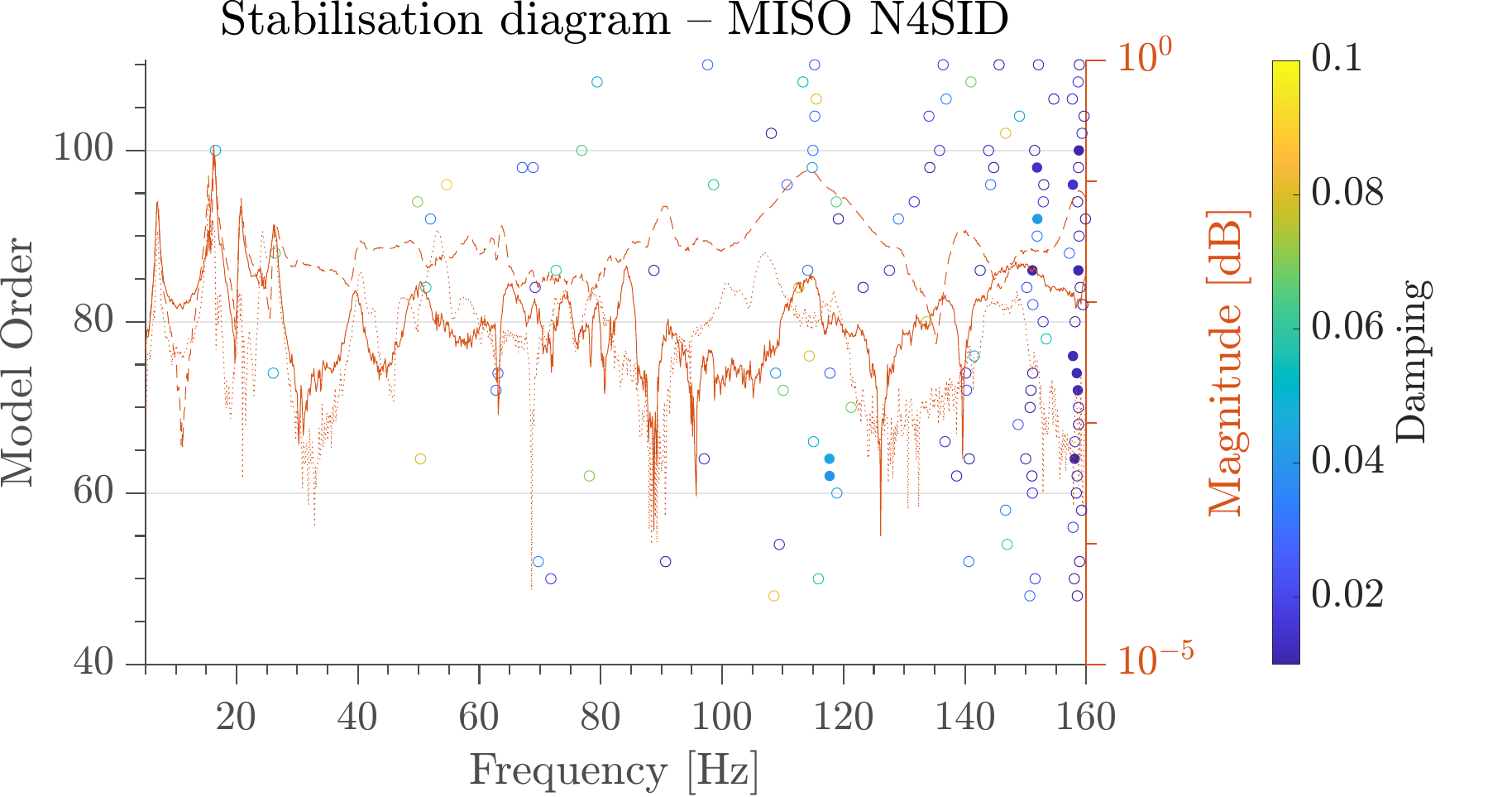}}
		\captionsetup{font={it},justification=centering}
		\subcaption{\label{fig:miso_n4sid}}	
	\end{subfigure}
    \begin{subfigure}[t]{.495\textwidth}
	\centering
		\includegraphics[width=\textwidth,keepaspectratio]{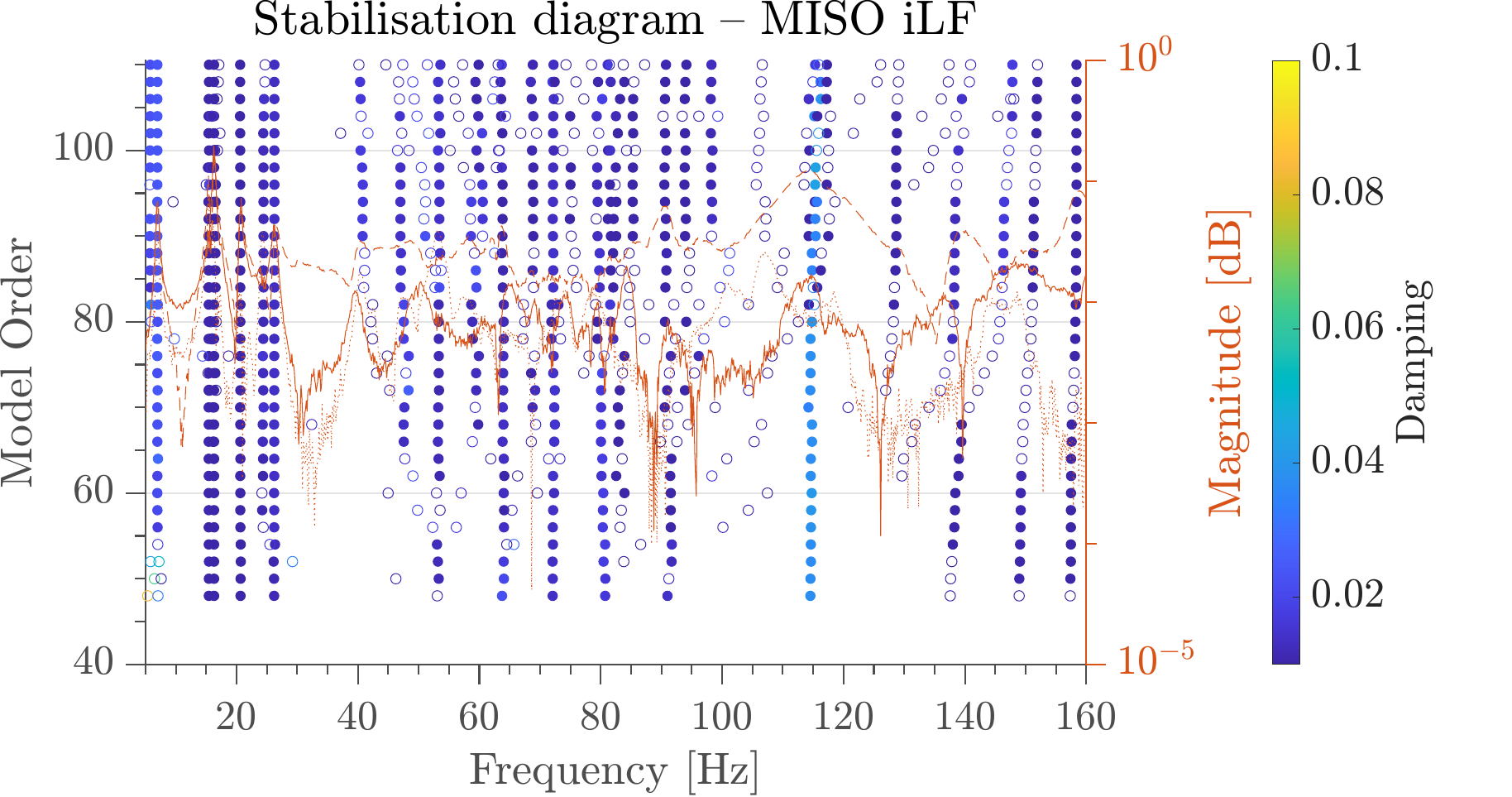}
		\captionsetup{font={it},justification=centering}
		\subcaption{\label{fig:miso_ilf}}	
	\end{subfigure}
	\caption{Hawk T1A Aircraft: Stabilisation diagrams for the MISO modal identification via N4SID and iLF.}
	\label{fig:miso_stab}
\end{figure*}

Notably, the SIMO and MISO identification is mostly unsuccessful with N4SID, as only one stable mode is identified in the MISO case, as confirmed in \cref{fig:miso_n4sid}, and none for the SIMO case. On the other hand, iLF (\cref{fig:miso_ilf}) can identify many new modes (8.7 \% more than the SISO), in particular in the 68-84 Hz interval, {as well as} most of the SISO modes. Even more, the MIMO iLF identification can detect even more modes as it considers all 91 channels  {located all} around the aircraft. This analysis identifies a total of 33 modes, five of them being at their first identification. To the authors' knowledge, this is the first time the MIMO modal identification of the BAE System Hawk T1A aircraft has been carried out.
In general terms, the $\omega_n$ identified by different cases are coherent and show little deviation, showing confidence in the proposed iLF identification approach.

\begin{table*}[!hbt]
\centering
\caption{Damping ratios identified by iLF, N4SID and RFP alongside the results in \cite{Haywood-Alexander2024} for the channels specified in \cref{fig:si_wing}. The values marked with (*) are those mistakenly reported in \cite{Haywood-Alexander2024} and corrected with the code provided in \cite{Champneys2023}.\label{tab:wing_damp}}
\resizebox{.75\textwidth}{!}{  
\begin{tabular}{lc c cc c cc cc cc}
    \hline
    \multicolumn{12}{c}{\textbf{Damping }[-]}\\\hline
    \textbf{Case} & \multicolumn{6}{c}{{SISO}} & \multicolumn{2}{c}{{SIMO}} & \multicolumn{2}{c}{{MISO}} & {MIMO} \\\cline{2-7}\cline{10-11}
    \textbf{Sensor}(s) & \multicolumn{3}{c}{{LTC-05}} & \multicolumn{3}{c}{{LLC-07}} & \multicolumn{2}{c}{{LTC-05 \& LLC-07}} & \multicolumn{2}{c}{{SW\_LC1}} & {All} \\\cline{2-4}\cline{8-9}\cline{12-12}
    \textbf{Mode} & {RFP} \cite{Haywood-Alexander2024} & {N4SID} & {iLF (10\textsuperscript{-4})} & {RFP} & {N4SID} & {iLF (10\textsuperscript{-4})} & {N4SID} & {iLF} & {N4SID} & {iLF} & {iLF} \\
    \hline
    \emph{1} & 0.0243 & 0.019 & 1.410 & 0.024 & 0.019 & 1.410 & - & 0.026 & - & 0.024 & 0.028 \\
    \emph{2} & 0.0155 & 0.005 & 1.197 & 0.014 & 0.005 & 1.197 & - & 0.012 & - & 0.006 & 0.008 \\
    \emph{3} & 0.0059 & 0.006 & 2.503 & 0.006 & 0.006 & 2.503 & - & 0.006 & - & 0.008 & 0.010 \\
    \emph{4} & - & - & - & - & - & - & - & - & - & - & 0.015 \\
    \emph{5} & - & - & - & - & - & - & - & 0.022 & - & 0.008 & 0.012 \\
    \emph{6} & 0.0085 & 0.013 & 0.038 & 0.009 & 0.013 & 0.038 & - & 0.013 & - & - & 0.012 \\
    \emph{7} & 0.0094 & 0.011 & 0.270 & 0.011 & 0.011 & 0.270 & - & 0.013 & - & 0.015 & 0.010 \\
    \emph{8} & 0.0158 & 0.015 & 0.212 & 0.016 & 0.015 & 0.212 & - & - & - & 0.013 & - \\
    \emph{9} & - & - & - & - & - & - & - & 0.015 & - & 0.013 & 0.016 \\
    \emph{10} & 0.0005 & 0.020 & 0.009 & 0.000 & 0.020 & 0.009 & - & 0.019 & - & - & - \\
    \emph{11} & 0.0001 & -0.028 & 0.013 & 0.000 & -0.028 & 0.013 & - & - & - & - & - \\
    \emph{12} & 0.0088 & 0.014 & 0.089 & 0.009 & 0.014 & 0.089 & - & 0.013 & - & 0.016 & 0.021 \\
    \emph{13} & 0.0072 & 0.007 & 0.014 & 0.015 & 0.007 & 0.014 & - & 0.007 & - & - & 0.016 \\
    \emph{14} & - & - & - & - & - & - & - & - & - & - & 0.017 \\
    \emph{15} & 0.0100 & 0.008 & 0.052 & 0.009 & 0.008 & 0.052 & - & 0.011 & - & - & 0.031 \\
    \emph{16} & 0.0047 & 0.007 & 0.033 & 0.011 & 0.007 & 0.033 & - & 0.009 & - & 0.015 & 0.024 \\
    \emph{17} & 0.0039 & -0.012 & 0.051 & 0.005 & -0.012 & 0.051 & - & 0.010 & - & 0.012 & - \\
    \emph{18} & 0.0045 & 0.006 & 0.137 & 0.005 & 0.006 & 0.137 & - & 0.005 & - & - & - \\
    \emph{19} & 0.0105 & 0.012 & 0.022 & 0.012 & 0.012 & 0.022 & - & 0.014 & - & 0.011 & - \\
    \emph{20} & 0.0050 & 0.018 & 0.407 & 0.017 & 0.018 & 0.407 & - & 0.010 & - & - & 0.006 \\
    \emph{21} & - & - & - & - & - & - & - & - & - & 0.013 & 0.013 \\
    \emph{22} & - & - & - & - & - & - & - & - & - & 0.014 & 0.019 \\
    \emph{23} & - & - & - & - & - & - & - & - & - & 0.011 & - \\
    \emph{24} & - & - & - & - & - & - & - & - & - & 0.013 & 0.030 \\
    \emph{25} & - & - & - & - & - & - & - & - & - & 0.004 & 0.016 \\
    \emph{26} & 0.0114 & 0.013 & 0.158 & 0.011 & 0.013 & 0.158 & - & 0.009 & - & 0.003 & - \\
    \emph{27} & 0.0280 & 0.003 & 0.004 & -0.021 & 0.003 & 0.004 & - & 0.010 & - & 0.011 & - \\
    \emph{28} & - & - & - & - & - & - & - & - & - & 0.008 & 0.006 \\
    \emph{29} & - & - & - & - & - & - & - & - & - & 0.013 & 0.025 \\
    \emph{30} & - & - & - & - & - & - & - & 0.006 & - & - & - \\
    \emph{31} & - & - & - & - & - & - & - & - & - & - & 0.021 \\
    \emph{32} & - & - & - & - & - & - & - & - & - & - & 0.017 \\
    \emph{33} & 0.0131 & 0.040 & 0.006 & 0.013 & 0.040 & 0.006 & - & 0.010 & - & 0.017 & 0.015 \\
    \emph{34} & 0.0034 & 0.048 & 0.013 & 0.004 & 0.048 & 0.013 & - & 0.032 & - & - & 0.015 \\
    \emph{35} & 0.0083 & 0.003 & 0.014 & 0.008 & 0.003 & 0.014 & - & 0.010 & - & - & 0.013 \\
    \emph{36} & - & - & - & - & - & - & - & - & - & 0.005 & - \\
    \emph{37} & - & - & - & - & - & - & - & - & - & 0.007 & - \\
    \emph{38} & 0.0124 (*) & 0.014 & 0.002 & 0.014 & 0.014 & 0.002 & - & 0.013 & - & 0.015 & 0.009 \\
    \emph{39} & - & - & - & - & - & - & - & 0.010 & - & - & - \\
    \emph{40} & - & - & - & - & - & - & - & 0.022 & - & 0.017 & 0.007 \\
    \emph{41} & - & - & - & - & - & - & - & - & - & - & 0.009 \\
    \emph{42} & 0.0151 & 0.019 & 0.006 & 0.015 & 0.018 & 0.031 & - & 0.014 & 0.010 & 0.007 & 0.010 \\
    \hline
\end{tabular}
}
\end{table*}

{Regarding} $\zeta_n$, the results for SISO identification with N4SID and iLF are mostly wrong and inaccurate. The former shows negative $\zeta_n$ (Modes \#11 and 17), and the latter negligible $\zeta_n$ values for all modes. This is expected, as N4SID and iLF are not meant to be used in a peak-picking fashion. Also, it should be noted that the RFP implementation in \cite{Haywood-Alexander2024,Champneys2023} shows negative damping for Mode \#27 of the LLC-07 sensor as well. Nevertheless, the SISO values obtained from RFP compare relatively well to those obtained by the SIMO and MISO identifications, validating the goodness of those identifications even more.

Next, the computational efficiency of N4SID and iLF on the experimental SIMO and MISO data should be examined. The N4SID identification of the SIMO case took 13 h 35 min, while iLF only 13.748 s {(a reduction of circa 99.97\% of the elapsed time)}. The model order for both methods was set between 46 and 110. For the MISO identification, with the same system and same model order range, the time to identification for N4SID was almost 21 h (20 h 57 min), while iLF obtained the results in 3.098 s {(circa 99.99\% reduction)}. The results were obtained in MATLAB 2021b running on a 14-core 2017 Apple iMac Pro with 128 GB of RAM. For reference, the same computation for the SIMO case with iLF takes 15.034 s (9.989 s for MISO) on a consumer-grade laptop (CPU 12th Gen Intel Core i5-1235U @ 1.30 GHz, 16 GB RAM, Windows 11 and MATLAB 2021b). 
Thus, the most complete identification of the system, the MIMO identification, is carried out solely with iLF due to the evident less computational efficiency of N4SID. The MIMO iLF identification takes 15.084 s on the iMac Pro and 10.710 s on the consumer-grade laptop.

\begin{figure}[!ht]
	\centering
		{\includegraphics[align=c,width=.7\textwidth,keepaspectratio]{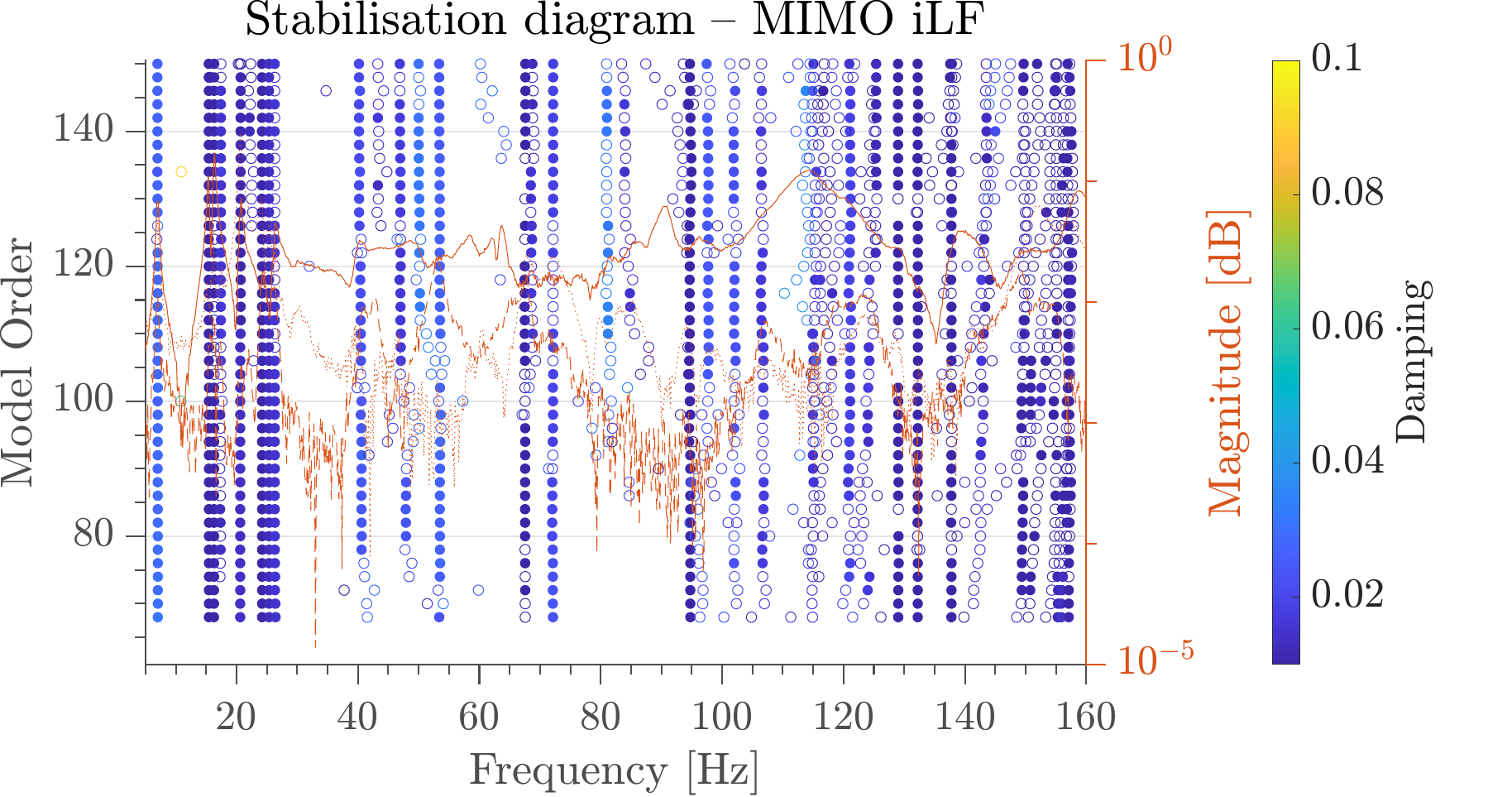}}
	\caption{Hawk T1A Aircraft: Stabilisation diagram of the MIMO test for all 91 accelerometer channels with iLF.}
	\label{fig:MIMO_stab}
\end{figure}

\Cref{fig:MIMO_stab} show the stabilisation diagram for the MIMO identification via iLF. This clearly shows the straight lines made up of the 33 stable modes.

\begin{figure}[!ht]
\centering
	\begin{subfigure}[t]{.35\textwidth}
	\centering
		{\includegraphics[width=\textwidth,keepaspectratio]{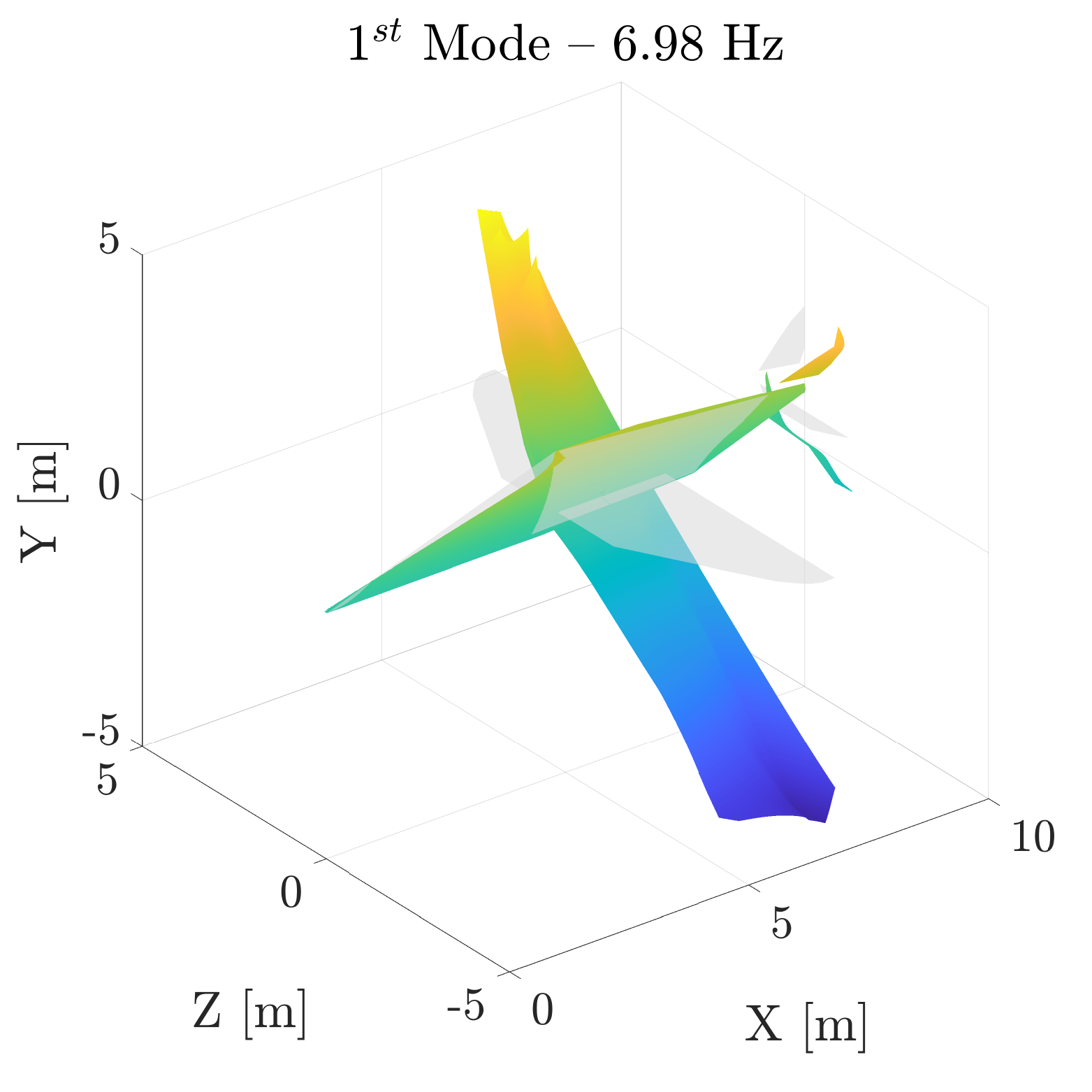}}
		\captionsetup{font={it},justification=centering}
		\subcaption{\label{fig:phi_1}}	
	\end{subfigure}
    \begin{subfigure}[t]{.35\textwidth}
	\centering
		\includegraphics[width=\textwidth,keepaspectratio]{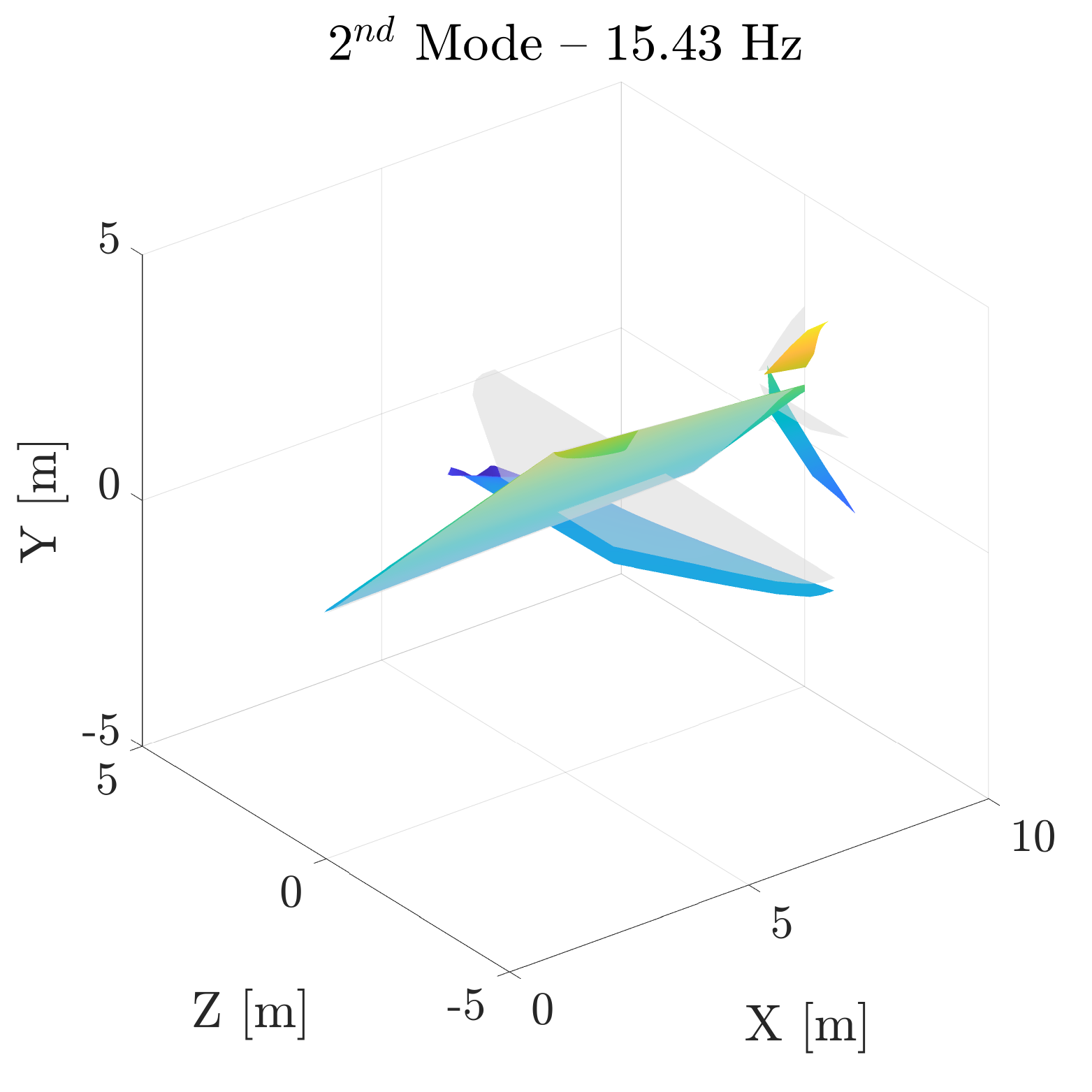}
		\captionsetup{font={it},justification=centering}
		\subcaption{\label{fig:phi_2}}	
	\end{subfigure}
 \begin{subfigure}[t]{.35\textwidth}
	\centering
		{\includegraphics[width=\textwidth,keepaspectratio]{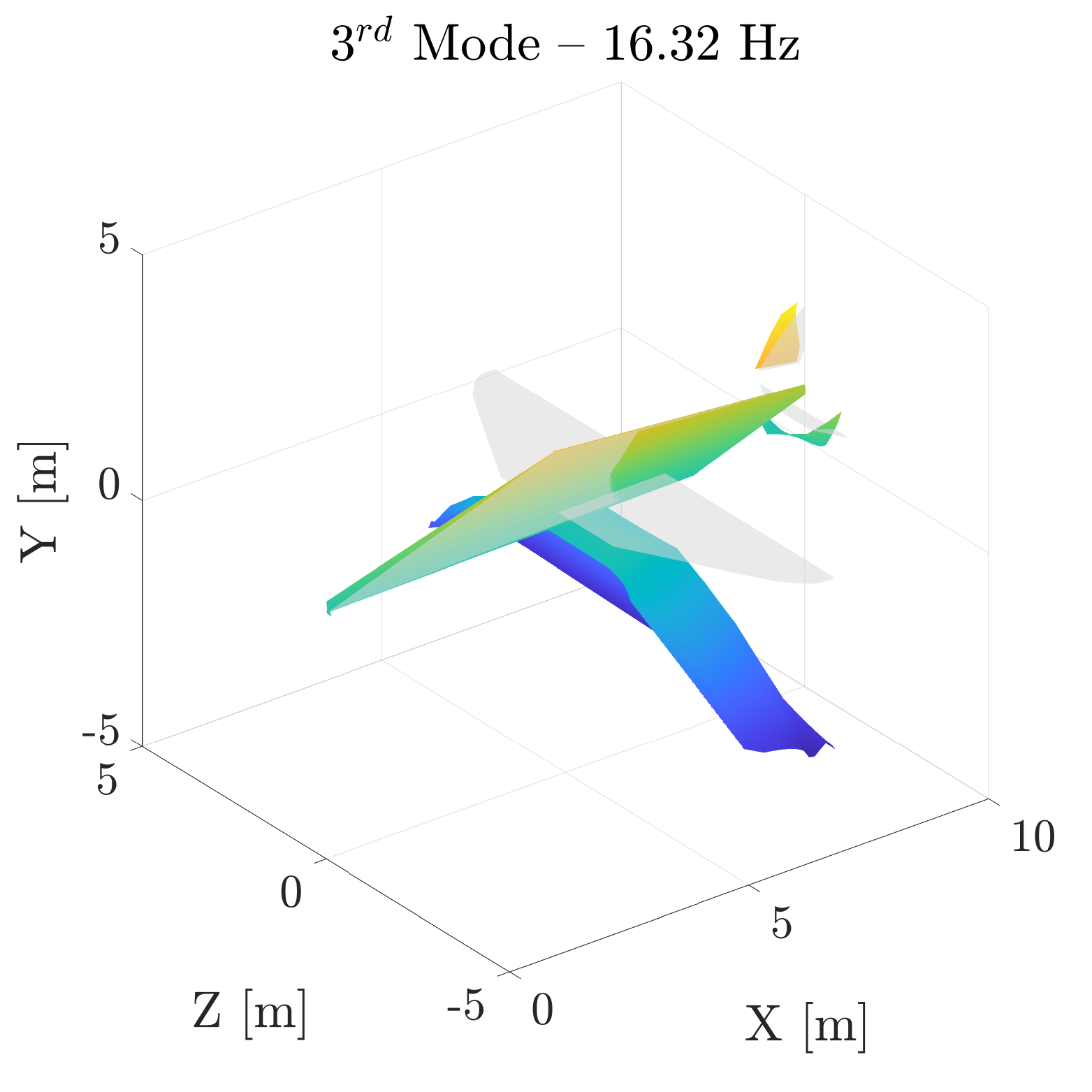}}
		\captionsetup{font={it},justification=centering}
		\subcaption{\label{fig:phi_3}}	
	\end{subfigure}
 \begin{subfigure}[t]{.35\textwidth}
	\centering
		{\includegraphics[width=\textwidth,keepaspectratio]{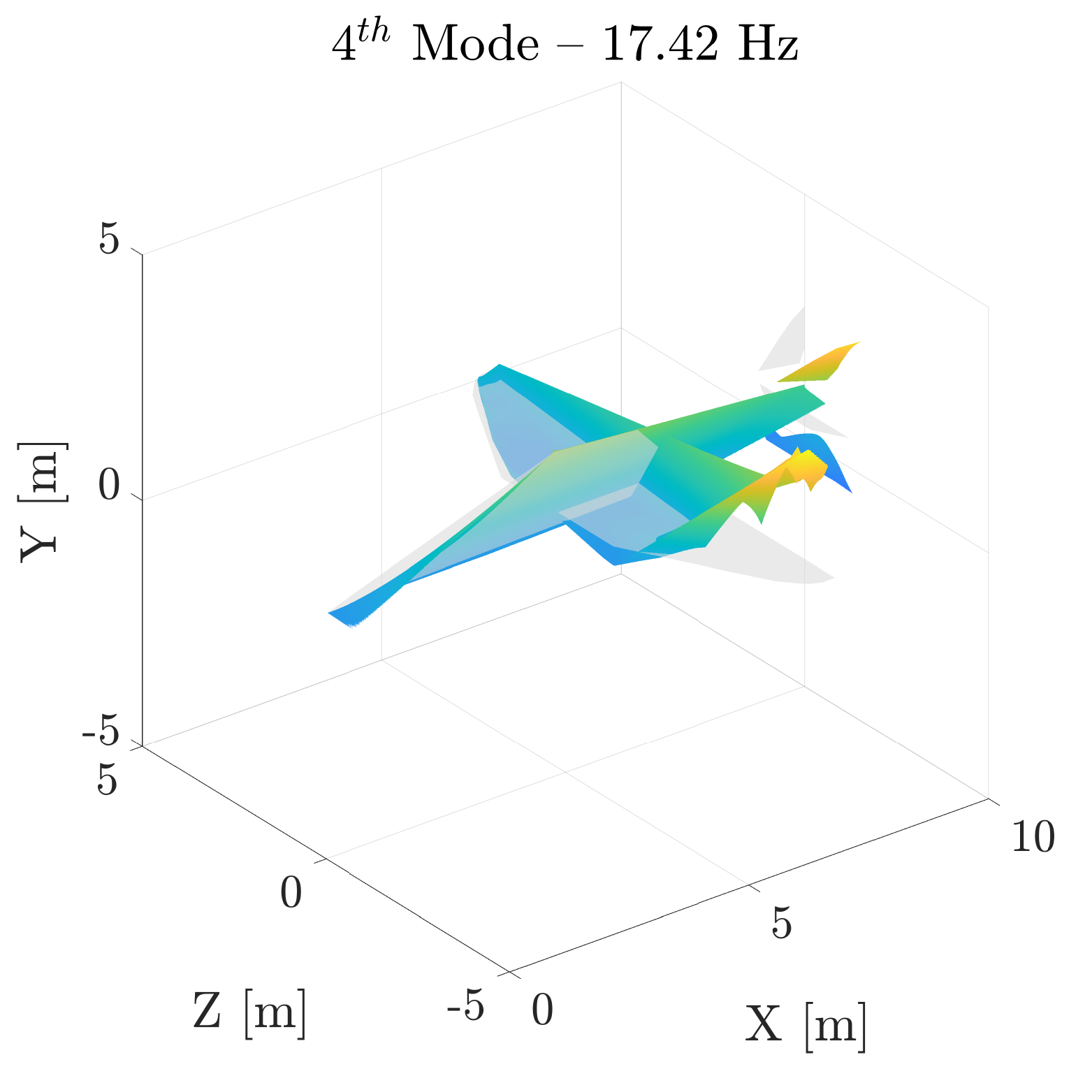}}
		\captionsetup{font={it},justification=centering}
		\subcaption{\label{fig:phi_4}}	
	\end{subfigure}
 \begin{subfigure}[t]{.35\textwidth}
	\centering
		{\includegraphics[width=\textwidth,keepaspectratio]{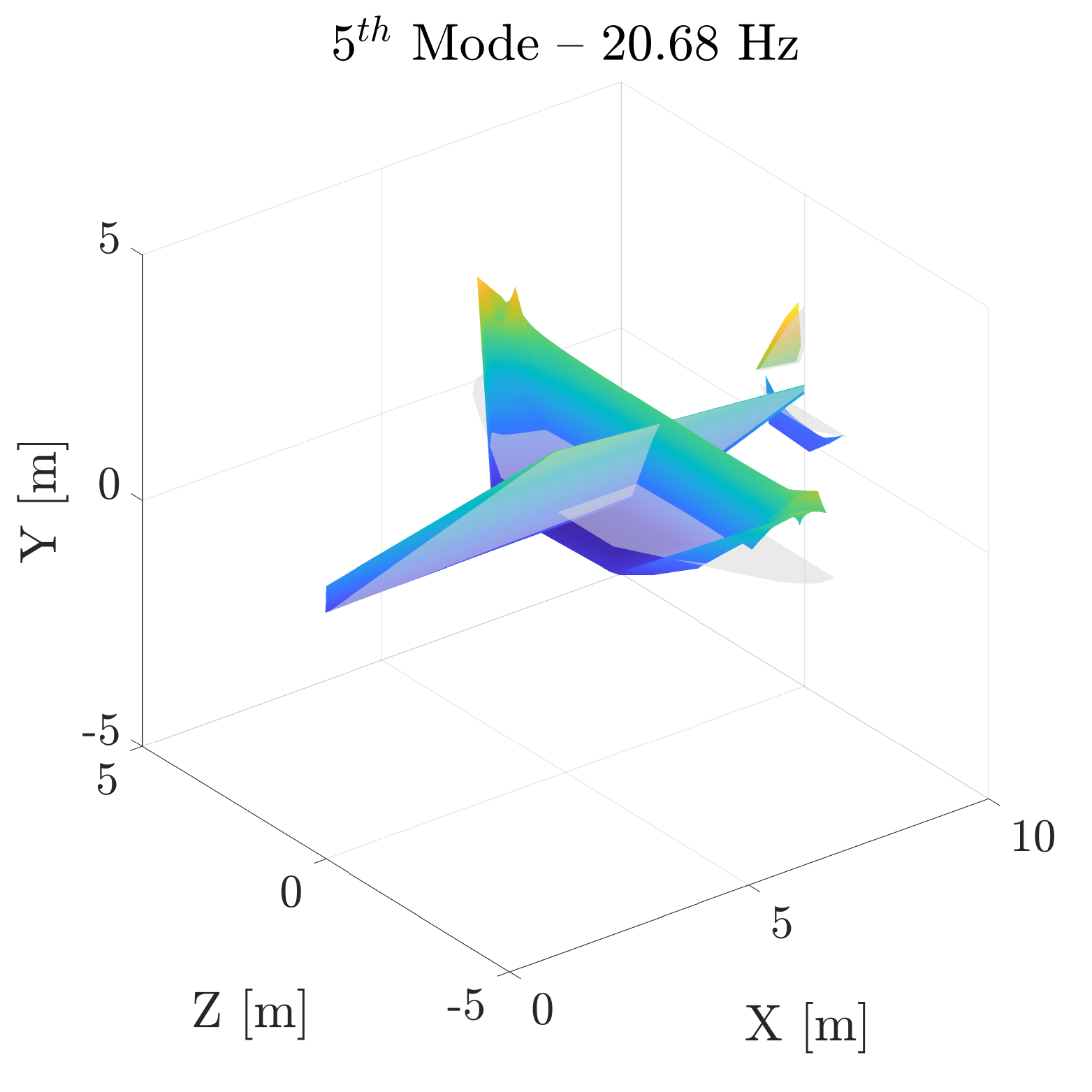}}
		\captionsetup{font={it},justification=centering}
		\subcaption{\label{fig:phi_5}}	
	\end{subfigure}
 \begin{subfigure}[t]{.35\textwidth}
	\centering
		{\includegraphics[width=\textwidth,keepaspectratio]{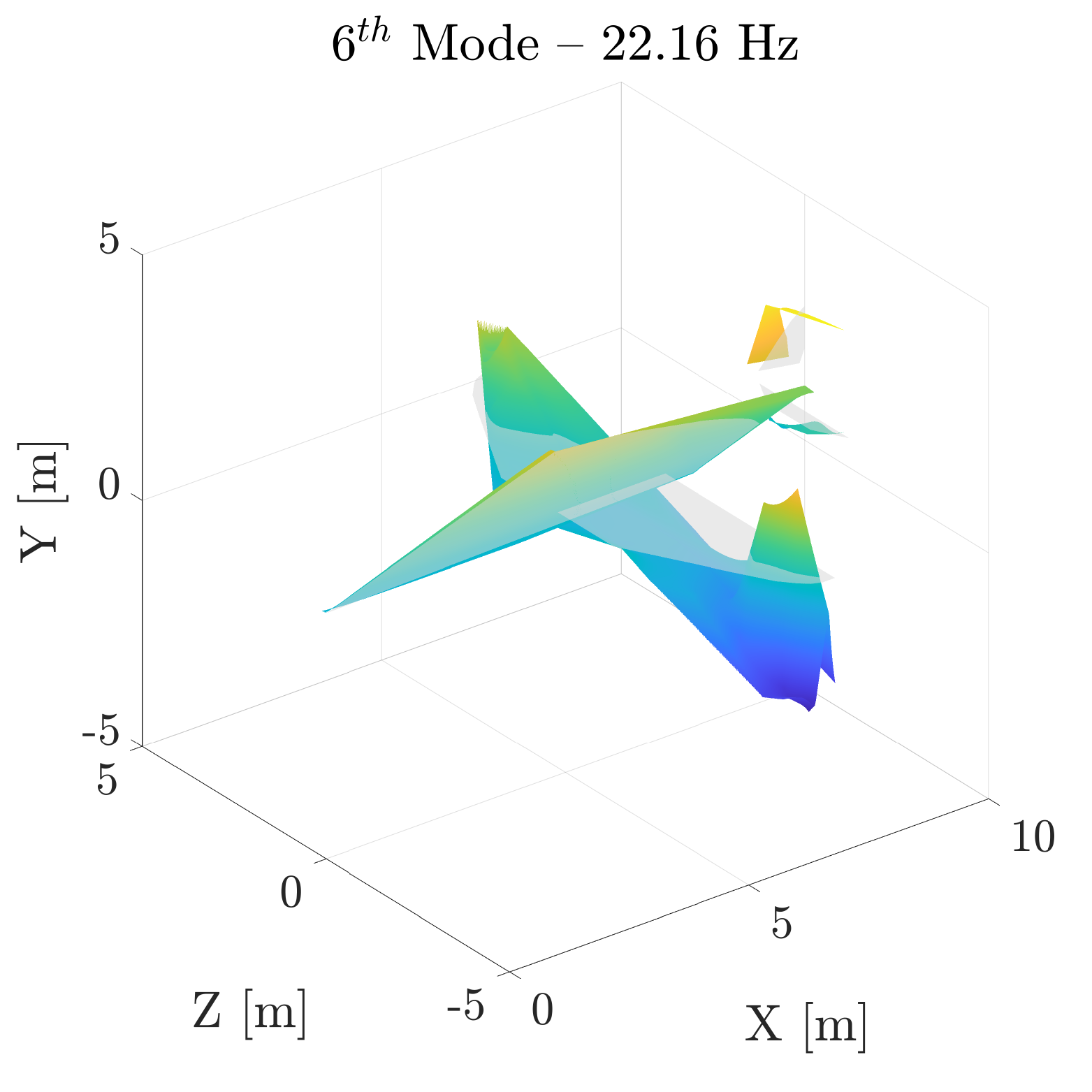}}
		\captionsetup{font={it},justification=centering}
		\subcaption{\label{fig:phi_6}}	
	\end{subfigure}
	\caption{Hawk T1A Aircraft: $\mathbf{\phi}_n$ of the first six modes.}
	\label{fig:phi_mimo}
\end{figure}

Having presented the identified $\omega_n$ and $\zeta_n$, the $\mathbf{\phi}_n$ identified from the MIMO dataset need to be discussed. These have been computed from all 91 accelerometer channels available; for conciseness, only the first six are shown in \cref{fig:phi_mimo}, and their explanation is given in \cref{tab:phi_mimo}.

For visualisation purposes only, the landing gear channels are not portrayed. The modal displacements have been scaled with respect to a characteristic length defined depending on the accelerometer position. The wing modal displacements are scaled to the aircraft semi-span (4.7 m), those relative to the fuselage to its maximum height (1.8 m), and those of the horizontal stabiliser displacements to its semi-span (2.18 m). Finally, the vertical stabiliser displacements are scaled to its span, 2.05 m. The only known size is the semi-span the others are retrieved from an existing CAD model \cite{Adams2022}.

It is clear that, due to the high number of channels some artifacts are created in the modal displacement. In particular, this is seen for $\mathbf{\phi}_{4,6}$ (\cref{fig:phi_4,fig:phi_6}), where the port wing tip measurement is quite unstable. It should be noted that some sensors are not aligned to the axis system defined in \cref{fig:MC_SENSORS_LOCATIONS2}, which means that suitable rotation matrices had to be generated to reconcile the displacements with the reference system directions. In addition, the signs of some displacements had to be corrected as many were {likely} placed facing down, with respect to the positive direction of the reference system.

\begin{table}[!hbt]
\centering
\caption{Hawk T1A Aircraft: Description of the first six modes  $\mathbf{\phi}_n$.\label{tab:phi_mimo}}
\begin{tabularx}{\textwidth}{lcX}
\hline
\textbf{Mode} & \multicolumn{1}{l}{\textbf{Natural Frequency {[}Hz{]}}} & \textbf{Definition}                                                                                                                                                                           \\ \hline
\textit{1}    & 6.98                                                    & Wing anti-symmetric (port side down) bending and torsion, horizontal stabiliser anti-symmetric (port side down) bending and torsion, and fuselage bending (port direction).                                     \\
\textit{2}    & 15.43                                                   & Dominated by horizontal stabiliser anti-symmetric (port side down) bending, with vertical stabiliser bending (port direction). Smaller contribution by wing symmetric downwards bending,                                                      \\
\textit{3}    & 16.32                                                   & Wing symmetric downwards bending and torsion, horizontal stabiliser symmetric upwards bending, vertical stabiliser bending (starboard direction), and fuselage bending (starboard direction). \\
\textit{4}    & 17.42                                                   & Wing symmetric upwards bending and torsion, horizontal stabiliser symmetric downwards bending, vertical stabiliser bending (port direction), and fuselage torsion.                            \\
\textit{5}    & 20.68                                                   & Wing symmetric upwards bending and torsion, horizontal stabiliser symmetric upwards bending, vertical stabiliser bending (starboard direction), and fuselage torsion                          \\
\textit{6}    & 22.16                                                   & Wing anti-symmetric (port side down) bending, horizontal stabiliser anti-symmetric (starboard down) bending and vertical stabiliser (starboard direction) bending.                                        \\ \hline
\end{tabularx}
\end{table}

The $\mathbf{\phi}_n$ appear more coupled than in other works, e.g. \cite{Kerschen2013}, due to the effect of the multi-input excitations and, in particular, of the specific setup. In fact, as remarked by the dataset authors in \cite{Wilson2024}, the same driving voltage is provided to all shakers, which, in structural terms, means exciting less stiff structures (e.g. the stabilisers) with the same force as stiffer parts, e.g. the main wings. This results in an energy concentration (also due to the close proximity of the three shakers acting on the stabilisers) in a less stiff area. This is also backed up by the $\mathbf{\phi}_{1-6}$, as they all show a significant contribution from the stabiliser displacements. All the displacement contributions for each mode are detailed in \cref{tab:phi_mimo}. All the raw data of the identified results are available in an accessible format in an open repository.

\section{Conclusions\label{sec:conc}}
This work introduces the computationally improved Loewner Framework (iLF) for the extraction of modal parameters (natural frequencies, damping ratios, and mode shapes) from multi-input multi-output (MIMO) data. Numerical and experimental validations have been carried out, yielding the following key findings:

\begin{itemize}
    \item {On the numerical dataset, the iLF is found to be as precise as N4SID and the standard LF for the extraction of modal parameters from multi-input multi-output (MIMO) data;}
    \item {On the same numerical dataset, the iLF identification from MIMO data is shown to be reliably robust to measurement noise, computationally less time-consuming than N4SID (2 orders of magnitude) and standard LF (an order of magnitude), and not affected by the presence of close-in-frequency modes as severely as LSCE is;}
    \item {On the experimental dataset, the iLF successfully extracts the modal parameters of trainer jet aircraft from MIMO data, also confirming its noise robustness and computational efficiency. Notably, to the best of the Authors' knowledge, this is the first instance in the literature that modal parameters are extracted from this very recent dataset.}
\end{itemize}

In summary, the iLF outperforms the said state-of-the-art methods, including its conventional counterpart, in terms of accuracy and computational performance. Hence, the authors suggest its use for the extraction of modal parameters, especially for MIMO test campaigns. {A guided tutorial for the iLF for modal analysis from MIMO data is shared in the Data Availability Statement, with all codes available to replicate the findings of this study.}

\appendix
\section{MATLAB Code Snippets of the Improvements}
\subsection{Code Vectorisation\label{sec:app}}
\noindent Original LF MATLAB implementation without vectorisation:
\begin{lstlisting}[language=Octave]
TL = eye(length(muRe));
TR = eye(length(laRe));   
for (i = 1:length(muIm)/2),  TL = blkdiag(TL,1/sqrt(2)*[1 1;-1j 1j]);  end
for (i = 1:length(laIm)/2),  TR = blkdiag(TR,1/sqrt(2)*[1 1j;1 -1j]);  end
% transform and discard possible small imaginary parts (due to roundoff)
LL = real(TL*LL*TR);  % discard possible imaginary parts
sLL = real(TL*sLL*TR); %  in the order of machine precision
% the left and right subsets in the new coordinate system
W = real(W*TR);
V = real(TL*V);
\end{lstlisting} 

\noindent LF implementation after vectorisation:
\begin{lstlisting}[language=Octave]
numPairsMu = length(muIm) / 2;
numPairsLa = length(laIm) / 2;
TL = eye(length(muRe) + 2 * numPairsMu);
TR = eye(length(laRe) + 2 * numPairsLa);
blockTL = 1/sqrt(2) * [1 1; -1j 1j];
blockTR = 1/sqrt(2) * [1 1j; 1 -1j];
idxMu = length(muRe) + 1 : length(muRe) + 2 * numPairsMu;
idxLa = length(laRe) + 1 : length(laRe) + 2 * numPairsLa;
TL(idxMu, idxMu) = kron(eye(numPairsMu), blockTL);
TR(idxLa, idxLa) = kron(eye(numPairsLa), blockTR);
LL = real(TL * LL * TR);
sLL = real(TL * sLL * TR);
W = real(W * TR);
V = real(TL * V);
\end{lstlisting}
\subsection{Code Decluttering\label{sec:app2}}
\noindent Original LF MATLAB implementation for the conversion from descriptor state-space to a continuous state-space model:
\begin{lstlisting}[language=Octave]
% A, B, C, D, and E are the matrices from the LF realisation
sis = dss(A,B,C,D,E); %create descriptor state space model
sys = ss(sis,'explicit'); %convert to continuous state-space model
A=sys.A;
\end{lstlisting} 

\noindent LF implementation after decluttering:
\begin{lstlisting}[language=Octave]
% A, B, C, D, and E are the matrices from the LF realisation
Am=A;Bm=B;Cm=C;Dm=D;Em=E;
A = pinv(Em)*Am;
\end{lstlisting}

{\small
\section*{{\small Corresponding Author}}
\noindent Gabriele Dessena \orcidlink{0000-0001-7394-9303} - Corresponding author\\
E-mail address: \href{mailto:gdessena@ing.uc3m.es}{gdessena@ing.uc3m.es}\\

\section*{{\small Author Contributions}}
\noindent {\small Conceptualisation, G.D. and M.C.; methodology, G.D. and M.C.; software, G.D.; validation, G.D. and M.C.; formal analysis, G.D.; investigation, G.D. and M.C.; resources,  G.D and M.C.; data curation, G.D.; writing---original draft preparation, G.D. and M.C.; writing---review and editing, G.D. and M.C.; visualisation, G.D.; funding acquisition, G.D and M.C..}

\section*{{\small Declaration of conflicting interests}}
\noindent {\small The author(s) declared no potential conflicts of interest with respect to the research, authorship, and/or publication of this article.}

\section*{{\small Funding}}
\noindent {\small
The first author disclosed receipt of the following financial support for the research, authorship, and/or publication of this article: This work has been supported by the Madrid Government (\emph{Comunidad de Madrid} - Spain) under the Multiannual Agreement with UC3M (\href{https://researchportal.uc3m.es/display/act564873}{IA\_aCTRl-CM-UC3M}).\\
The second author is supported by the \emph{Centro Nazionale per la Mobilità Sostenibile} (MOST -- Sustainable Mobility Center), Spoke 7 (Cooperative Connected and Automated Mobility and Smart Infrastructures), Work Package 4 (Resilience of Networks, Structural Health Monitoring and Asset Management).}}

\section*{{\small Acknowledgements}}
\noindent {\small
The datasets of the BAE Susyems T1A Hawk (\href{https://orda.shef.ac.uk/articles/dataset/BAE_T1A_Hawk_Starboard_Wing_Modal_Test/22710040}{BAE T1A Hawk Starboard Wing Modal Test} and \href{https://orda.shef.ac.uk/articles/dataset/BAE_T1A_Hawk_Full_Structure_Modal_Test/24948549/1}{BAE T1A Hawk Full Structure Modal Test}) have been retrieved from the University of Sheffield data repository ORDA and are cited in this work. The authors thank the datasets authors and the LVV at the University of Sheffield for making them openly available.
Furthermore, the authors would like to thank Prof Rauno Cavallaro of the Department of Aerospace Engineering at the Universidad Carlos III de Madrid for his valued advice.

\section*{{\small Data Availability Statement}\label{sec:6_data}}
{\small 
The improved Loewner Framework implementation (Data file 1) used in this work is openly available from the Zenodo repository at [\url{https://zenodo.org/records/13863292}] and the identification results and numerical dataset (Data file 2) supporting this study are openly available from the Zenodo repository at [\url{https://zenodo.org/records/13254981}].
This project contains the following underlying data:
\begin{itemize}
    \item Data file 1. A tutorial for the improved Loewner Framework for modal analysis;
    \item Data file 2. Data supporting: Improved tangential interpolation-based multi-input multi-output modal analysis of a full aircraft
\end{itemize}
In addition, this study used third-party experimental data (Data files 3 and 4) made available at [\url{https://orda.shef.ac.uk/articles/dataset/BAE_T1A_Hawk_Starboard_Wing_Modal_Test/22710040} and \url{https://orda.shef.ac.uk/articles/dataset/BAE_T1A_Hawk_Full_Structure_Modal_Test/24948549/1}] under licence that the authors do not have permission to share:
\begin{itemize}
    \item Data file 3. \href{https://orda.shef.ac.uk/articles/dataset/BAE_T1A_Hawk_Starboard_Wing_Modal_Test/22710040}{BAE T1A Hawk Starboard Wing Modal Test}
    \item Data file 4. \href{https://orda.shef.ac.uk/articles/dataset/BAE_T1A_Hawk_Full_Structure_Modal_Test/24948549/1}{BAE T1A Hawk Full Structure Modal Test}
\end{itemize}
Data files 1 and 2 are available under the terms of the [GNU General Public License v3.0 (GPL 3.0)].}

\bibliographystyle{elsarticle-num}

\end{document}